\documentclass[12pt]{iopart}

\usepackage{graphicx}
\usepackage{wrapfig}
\usepackage{appendix}
\usepackage{subfigure}
\usepackage{multirow}
\usepackage{bigdelim}

\begin{document}

\newcommand{\M}{\leavevmode\hbox{\kern3.1pt\tiny{M}\kern-8pt\footnotesize$\bigcirc$ }}
\newcommand{\xor}{\leavevmode\hbox{\footnotesize{XOR }}}
\newcommand{\nott}{\leavevmode\hbox{\footnotesize{NOT }}}
\newcommand{\cnot}{\leavevmode\hbox{\footnotesize{CNOT }}}

\newcommand{\ket}[1]{\ensuremath{|#1 \rangle}}
\newcommand{\bra}[1]{\ensuremath{\langle #1|}}
\newcommand{\braket}[2]{\ensuremath{\langle #1|#2 \rangle}}
\newcommand{\ketbra}[2]{\ensuremath{|#1 \rangle \langle #2|}}
\newcommand{\ro}[1]{\ensuremath{|#1 \rangle \langle #1|}}
\newcommand{\av}[1]{\ensuremath{\langle #1 \rangle}}

\newcommand{\real}{\ensuremath{\mathrm{Re}}}
\newcommand{\trace}{\ensuremath{\textsf{Tr}}}

\newcommand{\id}{\ensuremath{\mathsf{1}}}
\newcommand{\iden}{\ensuremath{{\sf 1\hspace*{-1.0ex}\rule{0.15ex}
{1.2ex}\hspace*{1.0ex}}}}
\newcommand{\R}{\ensuremath{{\sf R\hspace*{-0.9ex}\rule{0.15ex}
{1.5ex}\hspace*{0.9ex}}}}
\newcommand{\N}{\ensuremath{{\sf N\hspace*{-1.0ex}\rule{0.15ex}
{1.3ex}\hspace*{1.0ex}}}}
\newcommand{\Q}{\ensuremath{{\sf Q\hspace*{-1.1ex}\rule{0.15ex}
{1.5ex}\hspace*{1.1ex}}}}
\newcommand{\C}{\ensuremath{{\sf C\hspace*{-0.9ex}\rule{0.15ex}
{1.3ex}\hspace*{0.9ex}}}}

\newcommand{\h}[1]{\ensuremath{\mathcal{H}_{#1}}}

\newcommand{\me}{\ensuremath{\mathrm{e}}}
\newcommand{\mi}{\ensuremath{\mathrm{i}}}

\newcommand{\de}{\ensuremath{\mathrm{d}}}
\newcommand{\dd}[2]{\ensuremath{\frac{\mathrm{d}#1}{\mathrm{d}#2}}}
\newcommand{\ddd}[2]{\ensuremath{\frac{\mathrm{d}^2#1}{\mathrm{d}#2^2}}}

\newcommand{\ot}[2]{\ensuremath{\left( \begin{array}{c} #1 \\ #2
\end{array} \right)}}
\newcommand{\oth}[3]{\ensuremath{\left( \begin{array}{c} #1 \\ #2 \\ #3
\end{array} \right)}}
\newcommand{\twtw}[4]{\ensuremath{\left( \begin{array}{cc} #1 & #2 \\
#3 & #4 \end{array} \right)}}
\newcommand{\thth}[9]{\ensuremath{\left( \begin{array}{ccc} #1 & #2 & #3
\\ #4 & #5 & #6 \\ #7 & #8 & #9 \end{array} \right)}}

\newcommand{\expp}[1]{\ensuremath{\me^{\mi\hat{H}#1}}}
\newcommand{\expm}[1]{\ensuremath{\me^{-\mi\hat{H}#1}}}

\newcommand{\q}{\ensuremath{q_{ai}}}

\title{Surface code quantum computing by lattice surgery
}

\author{Clare Horsman$^1$, Austin G. Fowler$^2$, Simon Devitt$^3$ and Rodney Van Meter$^4$}
\address{$^1$ Keio University Shonan Fujisawa Campus, Fujisawa, Kanagawa 252-0882, Japan\\
$^2$ CQC2T, School of Physics, University of Melbourne, VIC 3010, Australia\\
$^3$ National Institute for Informatics, 2-1-2 Hitotsubashi, Chiyoda-ku, Tokyo 101-8430, Japan.\\
$^4$ Faculty of Environment and Information Studies, Keio University, Fujisawa, Kanagawa 252-0882, Japan} \ead{clare.horsman@gmail.com}

\begin{abstract}
In recent years, surface codes have become a leading method for quantum error correction in theoretical large scale computational and communications architecture designs.  Their comparatively high fault-tolerant thresholds and their natural 2-dimensional nearest neighbour (2DNN) structure make them an obvious choice for large scale designs in experimentally realistic systems.  While fundamentally based on the toric code of Kitaev, there are many variants, two of which are the planar- and defect- based codes.  Planar codes require fewer qubits to implement (for the same strength of error correction), but are restricted to encoding a single qubit of information.  Interactions between encoded qubits are achieved via transversal operations, thus destroying the inherent 2DNN nature of the code.  In this paper we introduce a new technique enabling the coupling of two planar codes without transversal operations, maintaining the 2DNN of the encoded computer.  Our lattice surgery technique comprises splitting and merging planar code surfaces, and enables us to perform universal quantum computation (including magic state injection) while removing the need for braided logic in a strictly 2DNN design, and hence reduces the overall qubit resources for logic operations. Those resources are further reduced by the use of a rotated lattice for the planar encoding. We show how lattice surgery allows us to distribute encoded GHZ states in a more direct (and overhead friendly) manner, and how a demonstration of an encoded CNOT between two distance 3 logical states is possible with 53 physical qubits, half of that required in any other known construction in 2D.
\end{abstract}
\maketitle
\section{Introduction}\label{intro}

Topological encoding of quantum data enables computation to be protected from the effects of decoherence on qubits and of physical device errors in processing. A logical qubit is encoded in the entangled state of many physical qubits; the exact ratio is determined by the \emph{code distance}, which is chosen based on measured physical error rates and desired logical error rates. As long as physical errors are below the \emph{threshold value}, increasing the number of physical qubits can exponentially suppress error on the logical qubits \cite{threshold}. Of the many types of codes known, the \emph{surface code} stands out as having the highest tolerance of component error ($\sim 1\%$ in recent results) when implemented on a simple 2-dimensional lattice of qubits with		
nearest-neighbour interactions \cite{topo-q-memory,raussendorfprl,austin1,austin14,new-threshold-austin,bombin,bomb11}. 

Surface codes were first introduced by Kitaev \cite{kitaev} in the context of anyonic quantum computing. There are two implementation strategies for such codes: the first uses exotic anyonic particles \cite{anyonrev}, and the second takes an active approach to error correction on a lattice of regular qubits. It is the latter that we are concerned with here. Within the active implementation, the first type of surface code that was developed was the \emph{planar code}: each logical qubit occupies a separate
code surface, with its own boundaries \cite{planar-bk,planar-fm}. While the error correction requires only nearest-neighbouring (NN) physical qubits to interact, multi-qubit gate operations were proposed to be performed transversally between surfaces. By contrast, the now standard surface code defines logical qubits as \emph{defects} (introduced degrees of freedom) within a single lattice, and deformation and braiding of the defects within a single surface performs gates between them. The price of maintaining NN interactions is over 3
times the number of physical qubits per logical qubit \cite{raussendorfprl,austin1}.

The reduced qubit requirements of the planar code make it very attractive for nearer-term experimental implementations of the surface code; for example the smallest correctable code (distance 3) uses 13 physical qubits for a single planar logical qubit, but a standard (double-defect) surface code uses 72. More broadly, smaller resource requirements are extremely useful for applications where small numbers of logical qubits need to be communicated in order to take part in distributed computing. However, the requirement for transversal two-qubit gates has previously made a planar encoding unfeasible for many systems where the physical qubits are confined in 2D and subject only to NN interactions, such as quantum dots \cite{qdos-architecture,optics-dots}, superconducting qubits \cite{super-archi,super-archi-austin}, trapped atoms \cite{atomic-archi}, nitrogen-vacancy (NV) diamond arrays \cite{2d-ss-architecture}, and some ion trap architectures \cite{2d-rf-iontraps,2d-penning-iontraps}. Until now, the only way to maintain NN interactions when performing multiple logical qubit gates was to move to a
defect-based surface code scheme.

In this paper we solve this problem by introducing a new method of deforming and combining planar code surfaces which we term \emph{lattice surgery}. By analogy with the term used in geometric topology, lattice surgery comprises the ``cutting" and ``stitching" of code surfaces to produce other planar surfaces. We show that these operations on the planar code produce novel code operations while requiring only standard NN physical interactions, and maintaining full fault-tolerance. Furthermore we demonstrate that these new operations can be combined to produce  multiple logical qubit gates without any transversal interactions, and we give the full construction for a \cnot operation between two planar qubits. To complete the universal gate set we demonstrate how magic states can be injected into the code space, and also give a useful direct construction of the Hadamard gate. We show how defect- and planar- based qubits can be interchanged, and demonstrate how a defect-based qubit can be detached from a code surface as a planar qubit. We finish by detailing two important medium-term achievable experiments that could be performed using lattice surgery: producing entangled planar qubits, both as Bell pairs and GHZ states; and a full NN \cnot between two distance-3 planar qubits. These use significantly fewer physical resources than defect-based codes, with our smallest lattice surgery \cnot requiring 53 qubits to implement: half the physical qubits of the smallest known defect-encoded \cnot operation, which we also describe.

\section{Surface codes}\label{scs}

\begin{figure}[t]
\centering
       \includegraphics[width=9cm]{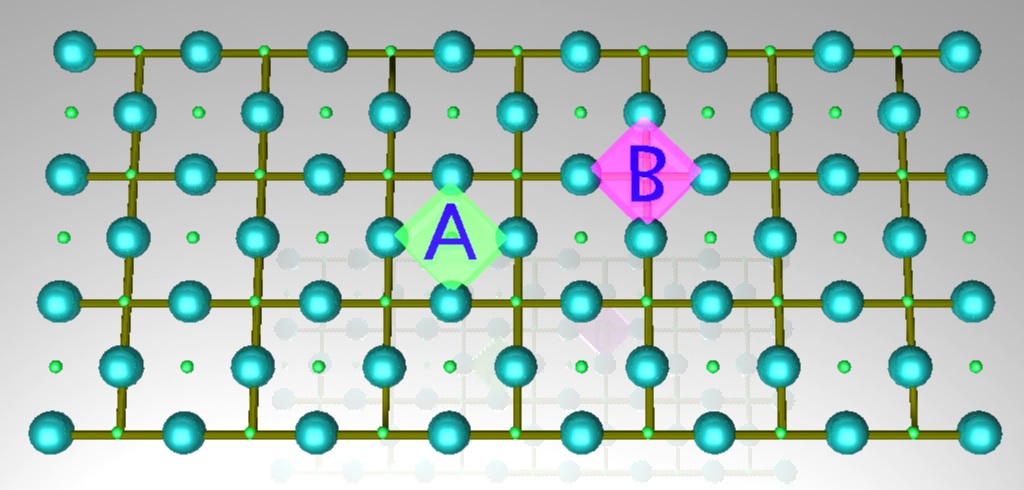}
	\caption{Part of the basic lattice of the surface code. Data qubits are shown large, syndrome qubits as small. The label `A' marks a face plaquette, `B' a vertex plaquette.}\label{plaquettes}
\end{figure}
\begin{figure}[t]
   \centering
    \subfigure[]
       {\includegraphics[width=6cm]{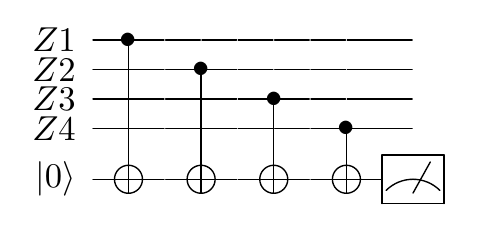}
        \label{zcn}}\hspace{1cm}
            \subfigure[]
       {\includegraphics[width=8cm]{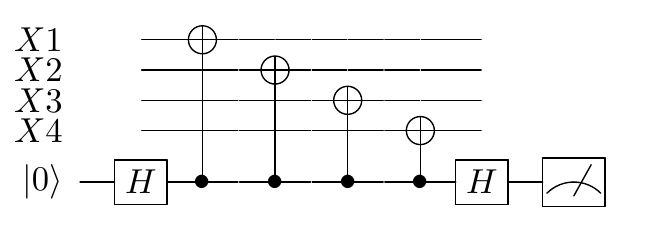}
        \label{xcn}}\hspace{1cm}
   \caption{Circuits for syndrome extraction: a) Z-syndrome b) X-syndrome. Measurements are in the computational basis.}
   \label{syndromes}
\end{figure}
Surface codes use a two-dimensional regular lattice of entangled physical qubits to give the substrate on which logical qubits are defined \cite{planar-bk,planar-fm}. The lattice is made up of two types of qubits, data and syndrome, differing only in their function within the code. Syndrome qubits are
repeatedly and frequently interacted with neighbouring data qubits and
measured to detect the presence of errors. Data qubits are measured
less frequently and only to perform computation. The  qubits are arranged in a lattice as in figure \ref{plaquettes}. The lines in the figure are aids to the eye, and do not designate any physical interactions or structures. The data qubits are in a simultaneous eigenstate of Pauli-$Z$ operators around each face (for example the ``plaquette" $A$ in figure \ref{plaquettes}), and Pauli-$X$ around each vertex (e.g.. $B$ in figure \ref{plaquettes}). That is, the \emph{stabilizers} of the systems are, for all face $F$ and vertex $V$ plaquettes,
\begin{equation}  \otimes_{i \in F} Z_i \ \ \ \ \mathrm{and} \ \ \ \ \  \otimes_{j \in V} X_j\label{lattstate}\end{equation}
\begin{figure}[t]
   \centering
    \subfigure[]
       {\includegraphics[width=5cm]{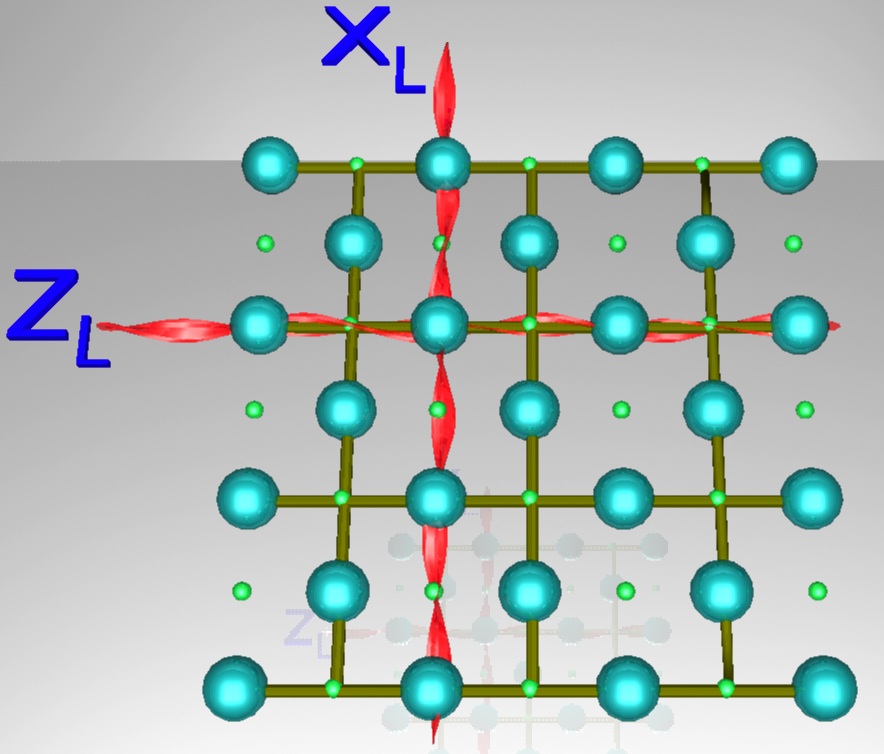}
       }\hspace{1cm}
            \subfigure[]
       {\includegraphics[width=5cm]{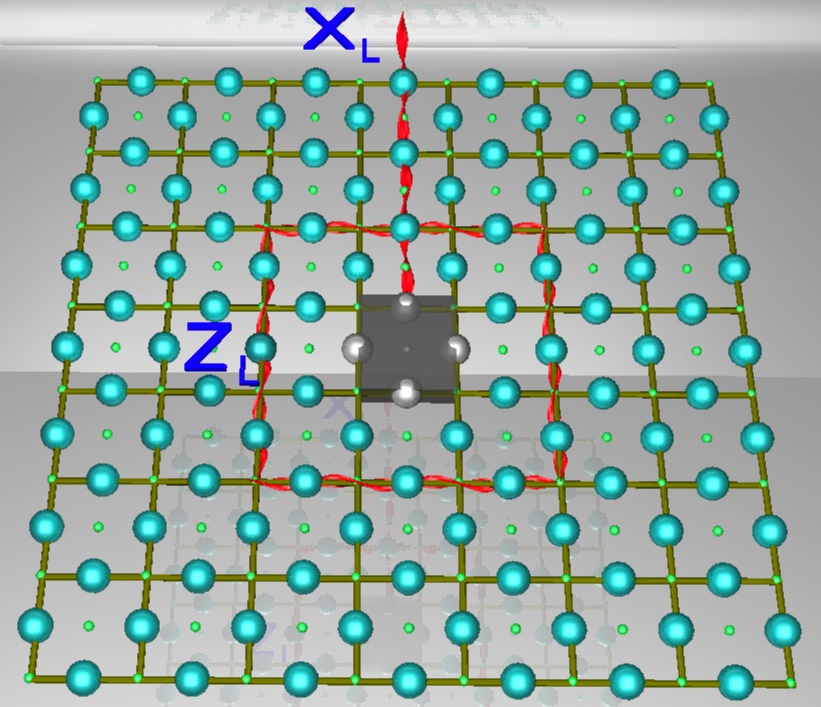}
        }\hspace{1cm}
        \subfigure[]
       {\includegraphics[width=8cm]{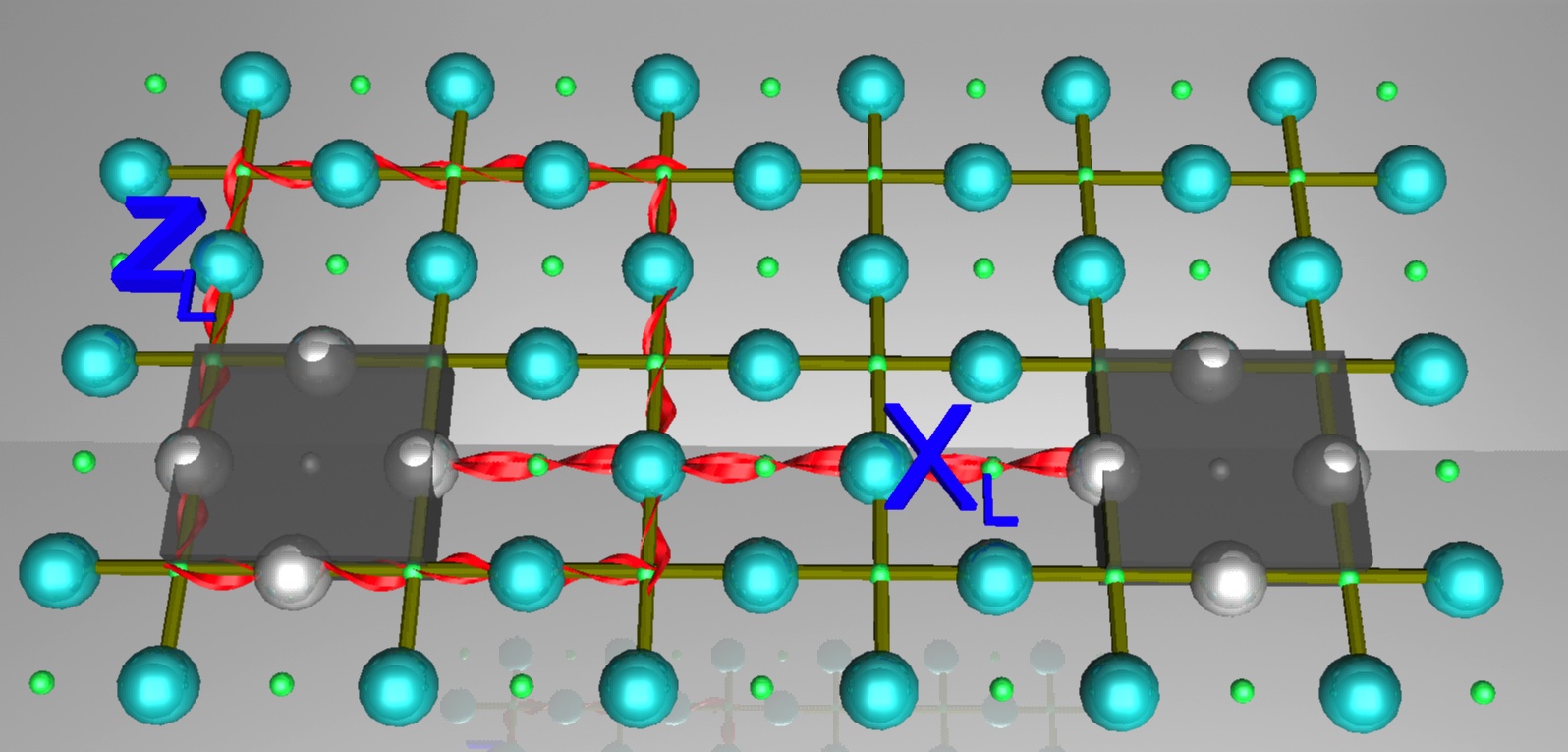}
       }\hspace{1cm}
   \caption{The three surface code methods of encoding a single logical qubit, shown here at distance 4: a) planar; b) single defect; c) double defect. Sample logical operators are marked in each case.}
   \label{distfour}
\end{figure}
\noindent where $\{i \in F\}$ and $\{j \in V\}$ are the data qubits in each face and vertex plaquette respectively.

The job of the syndrome qubits is to keep the lattice in a simultaneous eigenstate of these operators, in the face of possible qubit and measurement errors. To do this, syndrome qubits are placed in the centre of each plaquette, and in each round of error correction measure the 4-party stabilizer of that plaquette. The standard circuits for these \emph{syndrome measurements} are given in figure \ref{syndromes}. In the absence of the errors, the syndrome measurement will have
the same value in every round. Every time the syndrome measurement
changes value, an endpoint of a local chain of errors has been
detected. A pattern of chains of corrective operations highly likely
to maintain the correct logical state can be inferred using
Edmonds' minimum weight perfect matching algorithm \cite{Edmo65a,Edmo65b,Kolm09}.
Corrective operations are applied to classical data associated with
the measurement results and algorithm being executed, rather than the physical qubits. This ensures
that no corrective operations need to be applied to the qubits,
reducing the quantum error rate.

Qubits are defined using spare degrees of freedom in this lattice. In general, the requirement of the stabilizers (\ref{lattstate}) fixes the state fully; there are, however, several ways of introducing a degree of freedom into the lattice state. The first method is to define the lattice as having both \emph{rough} and \emph{smooth} boundaries as in figure \ref{distfour}(a). This introduces a single degree of freedom, and so the entire lattice can be used to encode a single logical qubit. This is the planar version of the surface code. A second method produces the required degree of freedom by not enforcing one of the stabilizers of equation (\ref{lattstate}) -- that is, introducing a \emph{defect} into the lattice, figure \ref{distfour}(b). In both cases we can define the logical operators of the encoded qubit, shown in the figures. In practice, in the defect-based code the qubits are defined by double defects, figure \ref{distfour}(c), as this localises the logical operators (no error chains can go between the defects and the edge of the lattice as they are opposite-type boundaries) and allows many qubits to be easily defined on a single surface. A third method involves introducing ``twists" to the lattice \cite{bombtwist}.

\begin{figure}[t]
\centering
       \includegraphics[width=8cm]{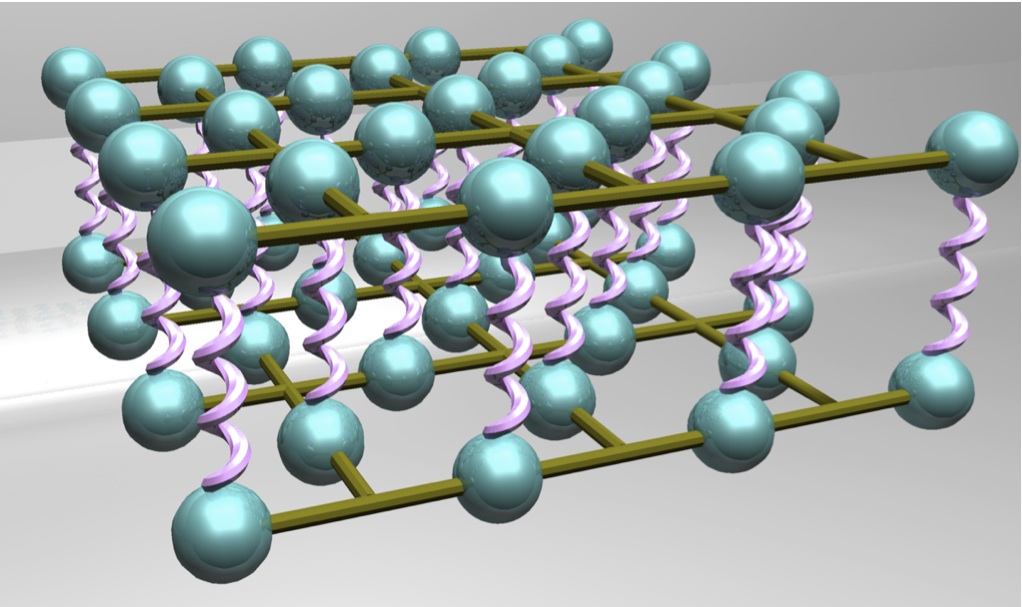}
	\caption{Transversal logical \cnot operation between two planar logical qubits. The pink interactions denote \cnot operations between pairs  of physical qubits. Syndrome qubits have been suppressed for clarity.}\label{transversal}
\end{figure}

In both planar and defect cases, a logical error is a chain of single-qubit errors that mimics a logical operator. Such error chains are undetectable, and cause a failure of the code. The \emph{code distance} is the measure of strength of the code, and is the length of the smallest undetectable error chain -- that is, the length of the smallest logical operator. For example, a code of distance 3 has the smallest logical operator as three physical qubit operations. Such a surface code can detect and correct a single physical error. Owing to their construction, a planar qubit of distance $d$ is generally around three times smaller than a defect-based qubit of the same distance. For both code types, $d$ rounds of error correction are needed fully to correct the lattice, in order to produce a spatio-temporal cell of depth $d$ to perform minimum weight matching of the ends of error chains \cite{raussendorf3D}. Whenever operations are performed on an encoded qubit, these $d$ rounds are required to correct the lattice before moving on.

Surfaces can initially be prepared in either the logical $\ket{0}$ or logical $\ket{+}$ state. To prepare a $\ket{0}_l$, all physical qubits are prepared in the $\ket{0}$ state, and then $d$ rounds of syndrome measurements performed to ensure fault-tolerance. Similarly, a $\ket{+}_L$ state is created by preparing all qubits in the physical $\ket{+}$ state and then performing syndrome measurements. At the end of the computation, the value of a logical qubit is read out by measuring all the qubits comprising the logical qubit in the measurement basis ($X$ or $Z$). These measurements are then subject to error correction, and the result of the logical measurement read from the parity of the logical operators measured.

The standard methods for performing two-qubit  gate operations (usually the \cnot operation) differ significantly between planar and defect-based surface codes. For a planar \cnot the original method was transversal: logical qubits are defined by construction on separate surfaces, and each physical qubit of one surface performs a \cnot with the corresponding physical qubit of the other surface, as shown in figure \ref{transversal}. After these operations, a \cnot has occurred between the logical qubits. By contrast, in the defect-based code there are no transversal operations, and the \cnot is performed by \emph{braiding} defects: extending a defect by measuring out qubits in a line, and passing this extended defect around the second logical qubit defect \cite{raussendorfprl}. The only operations other than measuring out individual qubits are those of the standard error correction procedure. A second method of performing such a gate is \emph{code deformation}: the boundaries of the code lattice itself are deformed around the defects, performing interactions on the logically encoded qubit \cite{bombin,bomb06,bomb11}.

The use of NN-only interactions for full computation has made the defect-based code the method of choice, as transversal operations create many more implementation problems than NN interactions. The new procedure of \emph{lattice surgery} that we introduce in this paper removes transversal two-qubit operations from the planar code, and not only allows a NN only \cnot operation to be performed, but also introduces new logical qubit operations (which we term ``split" and ``merge")  in a manner that is practical for a system that may ultimately be built

\section{Lattice surgery}\label{lsurgery}

The standard methods for implementing surface code gates non-transversally treat the lattice in ways familiar from algebraic topology \cite{algetop}, continuously deforming the lattice in order to achieve the required results. The methods proposed here break from this by introducing discontinuous deformations of the lattice, analogous to the operations of surface surgery in geometric topology (see for example \cite{mono}). We introduce the notions here of ``merging" and ``splitting" planar code lattices, and demonstrate the logical operations that they perform on the encoded data. 

Lattice merging occurs when two code surfaces become a single surface. This is implemented by measuring joint stabilizers across the boundaries of the surfaces during error correction cycles. Depending on which boundaries are joined, this operation behaves differently. Splitting of code surface is the opposite procedure, in which joint stabilizers are cut, forming extra boundaries that turn one code surface into two. Again, the types of boundaries created determine the exact nature of the final states. We will now demonstrate in detail the results of these four operations.

\subsection{Lattice merging}\label{lmerge}

\begin{figure}[t]
\centering
       \includegraphics[width=10cm]{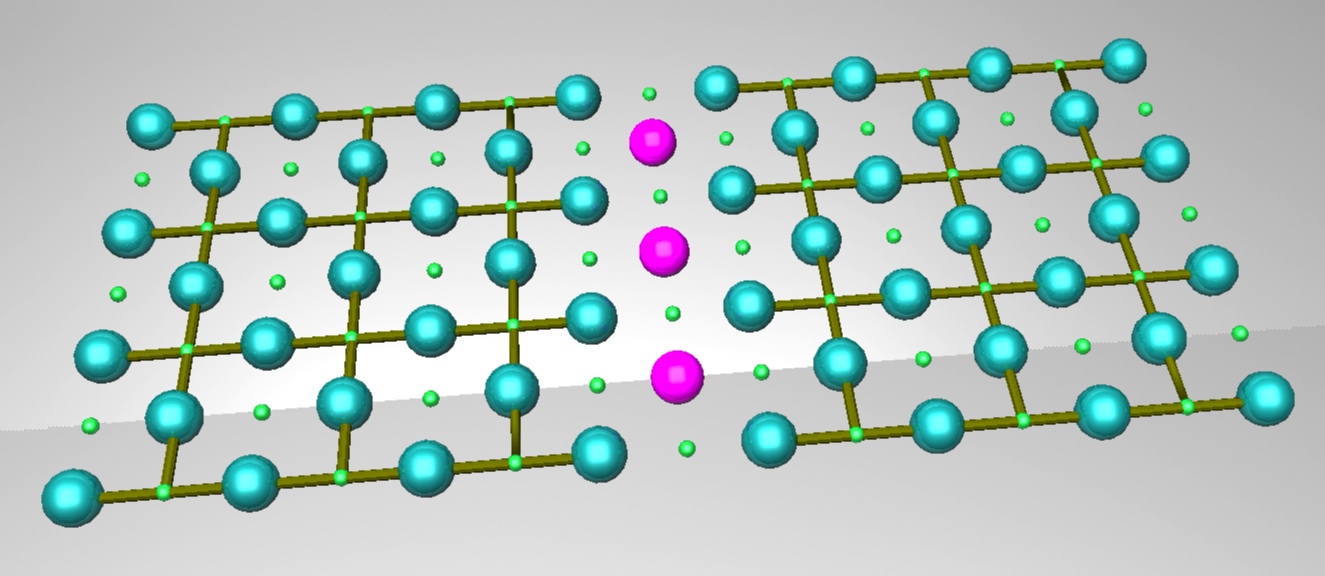}
	\caption{Arrangements of physical qubits for rough lattice merging. Left and right continuous surfaces encode separate logical qubits. The pink qubits form the intermediate qubit line for the merging operation.}\label{merge1}
\end{figure}

Let us consider the system shown in figure \ref{merge1}. There are two logical planar code surfaces, each separately stabilized and encoding a single logical qubit, and a row of `intermediate' uninitialised physical qubits. We merge the two systems by first preparing the intermediate data qubits in the state $\ket{0}$, and then performing $d$ rounds of error correction, treating the entire system as a single data surface.  

After correction, we will be able to reliably determine the sign of
the new $X$-stabilizer measurements spanning the old boundary. Note that
only the stabilizer measurements performed in the first round of the d
rounds will be known reliably, since only these are buried under
sufficient additional information to enable reliable correction. Later
rounds of stabilizer measurements become reliably known only as still
further rounds of error correction are performed.

Armed with reliable $X$-stabilizer measurements spanning the boundary,
we can reliably infer the eigenvalue of the product of these
$X$-stabilizers, which is equivalent to the two-logical-qubit operator
$X_LX_L$. The merge procedure described above is thus equivalent to
measuring $X_LX_L$. We define this merging operation as a rough merge:
the rough boundaries of the two surfaces are merged together. This
leads us to define a second type, that of a smooth merge, where it is
the smooth boundaries that are the subject of the merge operation. In
the case of a smooth merge, the intermediate qubits are prepared in
the $\ket{+}$ state before the new joint operators are measured. By
symmetry, a smooth merge is equivalent to measuring $Z_LZ_L$.

The lattice that remains now potentially has a sequence of syndrome measurements down the join that are incorrect. If there are an even number, then they can be corrected in the usual way by joining pairs with chains of $Z$ operations, as in \cite[\S V]{austin1}. In the case of an odd number of incorrect syndromes, the first one is not corrected, but simply tracked in software through the calculation (so that all subsequent correction operations correct to this ``incorrect" value). In this case we choose a particular $Z_L$ logical operator chain to be our ``reference" chain for that qubit; as long as the position of the chain is stored in memory and used for subsequent calculation and measurement, the logical qubit remains in the correct state. 

This action of measuring $X_LX_L$ on the two original qubits makes two things happen. Firstly, the state after measurement is non-deterministic, and correlated to the measurement outcome. Secondly, the planar surface now only has a single qubit degree of freedom, so we require a mapping from the original logical qubit states to the new logical qubit state post-merge. Let us consider the case where $\ket{\psi} = \alpha \ket{0}_L + \beta\ket{1}_L$ is merged with $\ket{\phi} = \alpha^\prime \ket{0}_L + \beta^\prime \ket{1}_L$. The outcome of merging is the measurement of $X_LX_L$; were this performed on two separate qubits then the state after the measurement is
\begin{equation} \frac{1}{\sqrt{2}}\left( \ket{\psi}\ket{\phi} + (-1)^M \ket{\bar{\psi}}\ket{\bar{\phi}} \right) \label{mergstate}\end{equation}
\noindent where $\ket{\bar{A}} = \sigma_x\ket{A}$, and $M$ is the outcome of the logical measurement, 0 or 1.

As is usual in surface code work, we now correct for the outcome of the measurement, leaving us with a deterministic state after the correction is applied. The correction will in practice be ``applied" by changing the interpretation of subsequent measurement outcomes, rather than by physical state operations. To see which corrections need applying, let us expand and re-write equation (\ref{mergstate}) dependent on the measurement outcome:
\begin{eqnarray}  (\alpha\alpha^\prime + \beta\beta^\prime)(\ket{00}_L + \ket{11}_L) +  (\alpha\beta^\prime + \beta\alpha^\prime)(\ket{01}_L + \ket{10}_L) & {\ \ \ \ \ \ } & M=0\nonumber \\
 (\alpha\alpha^\prime - \beta\beta^\prime)(\ket{00}_L - \ket{11}_L) +  (\alpha\beta^\prime - \beta\alpha^\prime)(\ket{01}_L - \ket{10}_L) & {\ \ \ \ \ \ } & M=1\label{ems}\end{eqnarray}

We also now need to consider the mapping to the new post-merge single qubit. A logical $\ket{0}_L$ state of the new surface will be the even-parity state of the combined $Z_L$ operator -- which is the simple product of the original $Z_L$ operators. The logical  $\ket{1}_L$ will be the odd-parity state of the product of the original $Z_L$ operators. However, we can see from equation (\ref{ems}) that the combinations of odd and even parity states that we have differ depending on the measurement result of the merge. We therefore have a conditional mapping, based on the $X_LX_L$ measurement outcome:
\begin{eqnarray}\ket{0}_L & \longrightarrow & \frac{1}{\sqrt{2}}(\ket{00}_L + (-1)^M \ket{11}_L) \nonumber \\
 \ket{1}_L & \longrightarrow & \frac{1}{\sqrt{2}}(\ket{01}_L + (-1)^M \ket{10}_L)\label{mnb} \end{eqnarray}
 
 If we now write the merge operation using the symbol ``$\M$", using equations (\ref{ems}) and this mapping, we find
 \begin{eqnarray}  \ket{\psi} \M \ket{\phi} & = & \alpha \ket{\phi} + (-1)^M \beta \ket{\bar{\phi}} \nonumber \\
 {} & = {} & \alpha^\prime \ket{\psi} + (-1)^M \beta^\prime \ket{\bar{\psi}}\label{stabrep}\end{eqnarray}

In classical terms, the truth table for the merge operation between qubits either in the state $\ket{0}_L$ or the state $\ket{1}_L$ is an \xor:

\begin{equation}
\centering
\begin{array}{c c | c}
\mathrm{In(1)} & \mathrm{In(2)} & \mathrm{Out}\\
\hline
0 & 0 & 0\\
0 & 1 & 1\\
1 & 0 & 1\\
1 & 1 & 0\end{array}
\end{equation}

 In the case of a smooth merge, the intermediate qubits are prepared in the $\ket{+}_L$ state before the new joint operators are measured. It is then the logical-$X$ operators that come merged, and so the action of the merge is an \xor in the Hadamard basis of the qubits:
\begin{eqnarray} \ket{\psi} \M \ket{\phi} & = & (a \ket{+}_L + b\ket{-}_L) \M ( a^\prime \ket{+}_L + b^\prime\ket{-}_L)\nonumber\\
{} & = & a^\prime\ket{\psi} + (-1)^Mb^\prime \ket{\bar{\psi}}\nonumber\\
{} & = &  a\ket{\phi} + (-1)^Mb\ket{\bar{\phi}}
\end{eqnarray}

 \ref{merge} illustrates the explicit transformations for two distance 2 codes. In the case of both smooth and rough merges, in order to preserve full fault-tolerance of the surface (and to gain a correct value for the logical $X_LX_L$ or $Z_LZ_L$ measurement), $d$ rounds of error correction are needed for a distance $d$ code to give the correct spatio-temporal volume for minimum weight matching of errors. Note that the distance of the merged surface in this configuration is the same as of the original surfaces: the length of the smallest error chain remains the same. Only if the merge increases the smallest error chain length does the code distance increase. This merge operation that we have defined has one very interesting property that sets it apart from other operations used on the surface code. While it is well-defined, fault-tolerant, and preserves the code space, it is not a unitary operation in the logical space. Two logical qubits are input into the operation, but only one emerges. 

\subsection{Lattice splitting}\label{lsplit}

The second type of new code operation, \emph{lattice splitting} is, in a sense, the converse operation to merging. A single logical qubit surface is split in half by a row of measurements that remove data qubits from the lattice. This leaves two separately stabilised surfaces at the end of the operation. As with merging there are two types, depending on the boundary along which the split occurs. 
\begin{figure}[t]
\centering
       \includegraphics[width=10cm]{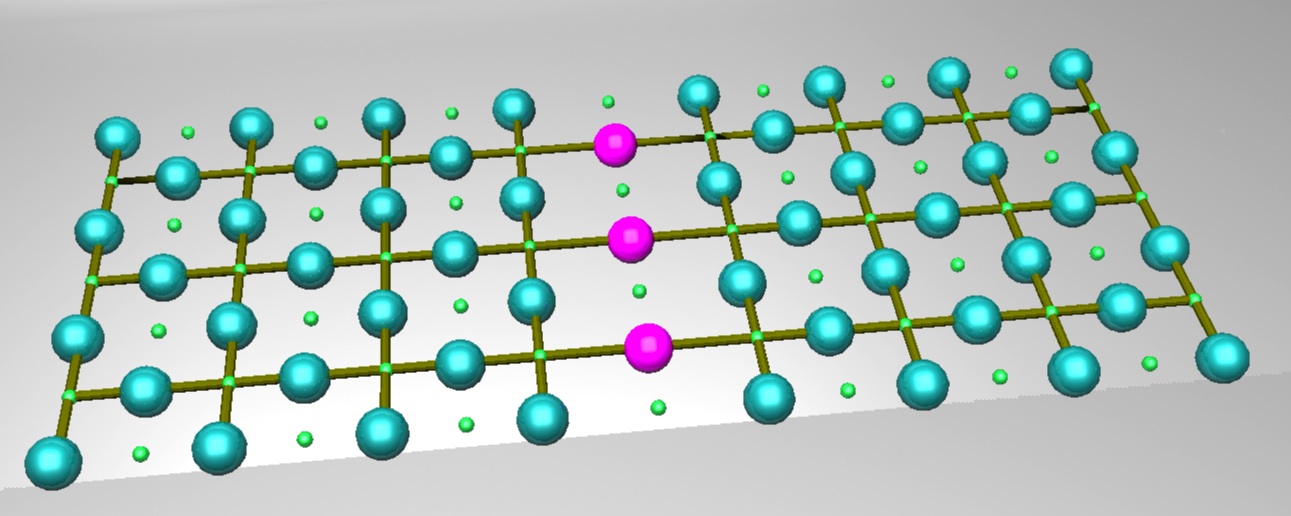}
	\caption{Arrangements of qubits for smooth lattice splitting. The pink qubits are measured out in the $X$-basis to leave two separately-stabilized logical qubit surfaces.}\label{split1}
\end{figure}

Let us consider first the \emph{smooth split}, shown in figure \ref{split1}. The middle row of qubits, shown, is measured out in the Pauli-$X$ basis. This is the same configuration as for a smooth merge, in which case the marked qubits would be the intermediate qubits, initialised in the $\ket{+}$ state. As with the merge, we can see the action of the split operation through the effect on the
logical operators.

In the case of the smooth split, after measuring out the intermediate qubits we are left with two surfaces that are then individually stabilized, as before for a total of $d$ rounds of error correction, where $d$ is the code distance. Unlike in the merge case, splitting can change the code distance: a split that divides a square surface symmetrically will halve the code distance. To end up with two surfaces of distance $d$, then, we need to start with one surface of size $d\times 2d$ (which also has code distance $d$).  After the split has been performed, we can look at the new plaquette operators on the join. Firstly, we can see that none of the joint $Z$ operators change at all: measuring out the qubits removes a row of face plaquettes from the error correction entirely, and leaves the surrounding face plaquettes untouched. The states of all three qubits before and after the split are therefore in the same superposition of eigenstates of the $Z$ logical operator.

The action of the split on the $X$ logical operator is more complicated. The set of $X$-measurements on the row of qubits will each have a random outcome, 0 or 1. Each measurement leaves 3-qubit $XXX$ plaquettes on either side of split, and the parity that then needs to be tracked for the purposes of error correction is the product of the measurement of this stabilizer with the measurement outcome at the split of what was the 4th qubit in the plaquette. The parity of the logical $X$ operator of the state of the remainder of the surface nevertheless remains fixed. However, as we now have two separate surfaces, this logical state is distributed across the two surfaces: the two surfaces are defined by reference to a single joint logical operator, rather than their individual ones. That is, the two surfaces will be in an entangled state of their logical $X$ operators. 

It is important to note that the action of the split means that, for the individual surfaces, there is no longer full equivalence between each set of $X_i$ operators across the surface that can make up an $X_L$ operator. This can be dealt with in two ways. First, a single line of $X_i$ operators is defined to be the $X_L$ operator (for example, individual $X$ operators on the top row of qubits). As long as this definition is the same on the two surfaces, subsequent computational use of these qubits will preserve their state. The second alternative is to correct the surface based on the measurement results so that all putative $X_L$ operators again become equivalent. This may be done by ``pairing" those 3-term stabilizers on the split that have negative parity, as in the merging. This may be performed either by physically performing the pairing operations (in practise not an efficient method as additional errors may be introduced), or by keeping track of these operations in the interpretation of future measurement results.

We can therefore write out the action of a smooth split, which preserves the logical $Z$ operator but splits the logical $X$ over the two qubits produced, once the necessary corrections or logical operator definitions have been arranged:
\begin{equation} \alpha\ket{0}_L + \beta \ket{1}_L \longrightarrow \alpha\ket{00}_L +  \beta \ket{11}_L\label{stabsplit}\end{equation}

By exchanging $X$ and $Z$ in the above argument, we can find the result of a rough split, which preserves the logical $X$ operator but splits the logical $Z$:
\begin{equation} a\ket{+} + b \ket{-} \longrightarrow a\ket{++} + b \ket{--}\end{equation}

As with the merge operation, the split operation is not unitary as the number of qubits has not been preserved. However, unlike the case of a merge, information has not been lost at the logical level: the original state on a single qubit can be recovered logically by performing a reversing merge operation after the split. Distributing a state across two logical qubits as the split does is indeed a common feature of surface codes. This is exactly what is done when a double defect is used to encode a single logical qubit: in fact what is encoded is an entangled pair, which is itself an encoding of a single qubit.  \ref{split} illustrates the explicit transformations for two distance 2 codes.

\section{Universal gate operations with lattice surgery}\label{universal}

We have now defined the operations of lattice surgery, splitting and merging the code surfaces. What is not immediately clear from the definitions is whether these operations are universal for quantum computing in the logical space of the planar surface code. We now demonstrate that this is in fact the case, by constructing a standard universal set comprising a logical \cnot gate and arbitrary logical single-qubit rotations (using magic state distillation and injection, as in the case of the standard surface code). We also give a method for performing the Hadamard gate as a basic code operation.

\subsection{The \cnot gate}\label{cnot}

\begin{figure}[t]
\centering
       \includegraphics[width=9cm]{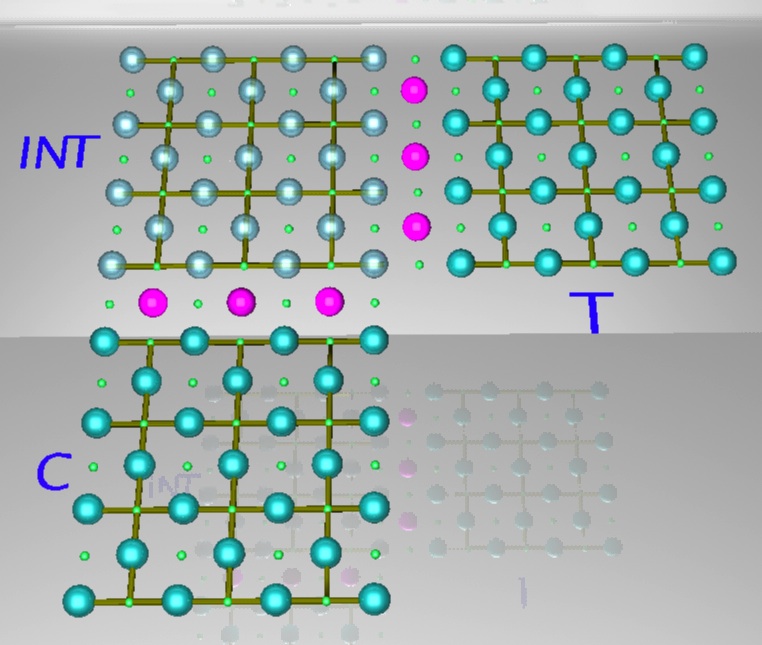}
	\caption{Layout of qubits for a \cnot operation with lattice surgery. Control (C) and target (T) surfaces interact by merging and splitting with the intermediate surface (INT).}\label{unicnot}
\end{figure}

The construction of a full \cnot gate using lattice surgery is shown in figure \ref{unicnot}. We start with the two logical qubits of distance $d$ that are the control, in state 
\begin{equation} \ket{C} =\alpha\ket{0} + \beta\ket{1}  = \bar{\alpha}\ket{+}_L + \bar{\beta}\ket{-}_L \end{equation}

\noindent and the target, in state 
\begin{equation} \ket{T} = \alpha^\prime\ket{0}_L + \beta^\prime\ket{1}_L \end{equation}

 \noindent We also have an intermediate logical qubit surface of distance $d$ initialised to the logical $\ket{INT} = \ket{+}_L$ state, and two strips of physical qubits, the first all in the physical $\ket{+}$ state and the second all in the physical $\ket{0}$ state, to help with merge operations.

The first step is to smooth merge the surfaces $C$ and $INT$. After the conditional definition of logical state post-merge, this creates a single surface in the state
\begin{eqnarray} \ket{C} \M \ket{INT} &  = & \bar{\alpha}\ket{+}_L + (-1)^M \bar{\beta}\ket{+}_L \nonumber\\
{} & = & \frac{1}{\sqrt{2}}\Big( \bar{\alpha} +  (-1)^M \bar{\beta}\Big) \ket{0} +  \frac{1}{\sqrt{2}}\Big( \bar{\alpha} -  (-1)^M \bar{\beta}\Big) \ket{1} 
\end{eqnarray}

\noindent where, as before, $M$ is the measurement outcome of the $Z_LZ_L$ operator performed during the merge. If $M$ is even then we redefine the basis of this qubit by a bit-flip (this need not be performed, but simply tracked in software). We then have
\begin{equation} \ket{C} \M \ket{INT} = \alpha\ket{0} + \beta \ket{1}\end{equation}

For full fault-tolerance, $d$ rounds of error correction are performed to create this merge operation. The code distance after the merge is still $d$. We then split this new surface back into the original two surfaces with a smooth split, measuring out the qubits along the split in the $X$ basis. With the logical-$X_L$ operators either redefined or the surface corrected, the state of these two new surfaces is then
\begin{equation}  \ket{C^\prime \ INT^\prime} = \alpha\ket{00}_L + \beta\ket{11}_L\end{equation}

\noindent The code distance of both qubits is again $d$, and $d$ rounds of error correction are required for this step as well. We now perform a rough merge operation between the surfaces $INT$ and $T$:
\begin{eqnarray} \ket{C^\prime \ (INT^\prime \M T) }& = & \alpha\ket{0}_L\otimes ( \ket{0}_L \M \ket{T} ) + \beta \ket{1}_L \otimes(\ket{1}_L \M \ket{T})\nonumber\\ 
{} & = &{}  \alpha\ket{0}_L\otimes \ket{T}  + (-1)^{(M^\prime)} \beta \ket{1}_L \otimes \ket{\bar{T}}\end{eqnarray}

\noindent where $M^\prime$ is again the merge measurement outcome, and again we have redefined the logical states dependent on the merge measurement outcome. This operation also takes $d$ rounds of error correction, and at the end we are left with two logical qubits of distance $d$. The outcome of this operation, as can be seen from the above equation, is a fully reversible \cnot operation. Fault tolerance has been maintained throughout with the multiple rounds of error correction, and the fact that at no point during this operation has the code distance of any of the logical surfaces used dropped below $d$. We have therefore constructed a fault-tolerant, unitary \cnot operation from lattice surgery operations without transversal gates.

\subsection{State injection}\label{stateinjection}

The second element that we need for a universal gate set is state injection. This allows for both state preparation and also arbitrary single-qubit rotations, in accordance with standard techniques  \cite{magic-dist}. As with all surface code implementations, rotations on the code surface cannot in general be performed by manipulating only the code surface of the single qubit. Arbitrary rotations require the use of ancilla states, with \cnot operations between the ancilla and logical qubit in order to implement a rotation gate on the logical qubit. 

\begin{figure}[t]
   \centering
    \subfigure[]
       {\includegraphics[width=4cm]{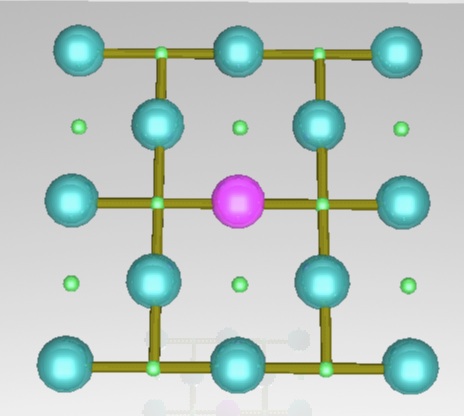}
        }\hspace{1cm}
            \subfigure[]
       {\includegraphics[width=4cm]{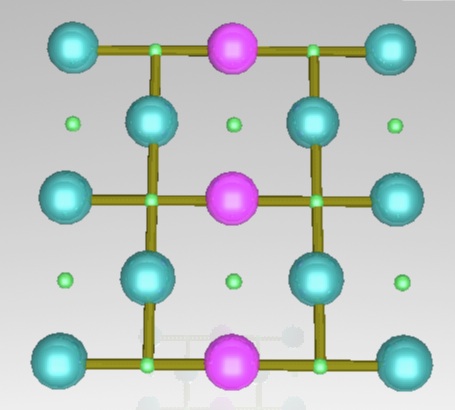}
         }\hspace{1cm}
                   \subfigure[]
       {\includegraphics[width=4cm]{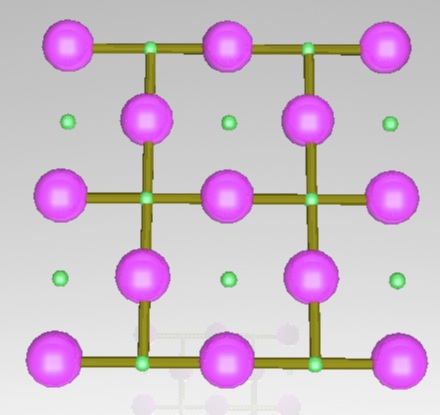}
        }\hspace{1cm}
            \subfigure[]
       {\includegraphics[width=4cm]{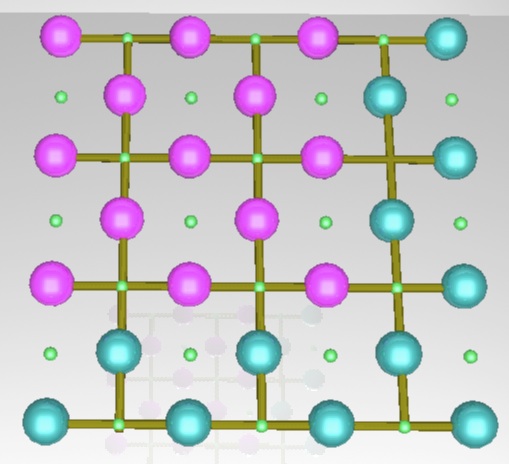}
         }\hspace{1cm}
            \subfigure[]
       {\includegraphics[width=4cm]{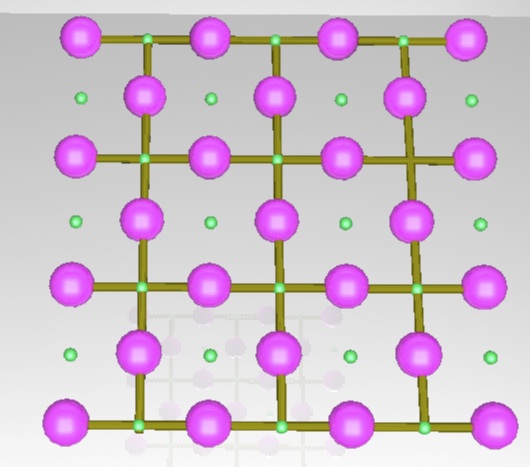}
         }
   \caption{Injecting an arbitrary state $\ket{\Psi} = \alpha \ket{0} + \beta \ket{1}$ into a planar surface: a) the blue qubits are prepared in \ket{0}, and the pink is the entangled state $\ket{\Psi}$; b)  \cnot operations are performed to create $\alpha \ket{000} + \beta\ket{111}$ on the pink qubits; c) measuring stabilizers gives a distance-3 surface in the logical $\ket{\Psi}$ state; d) prepare the blue qubits in \ket{0}; e) merging in the blue qubits gives a distance-4 surface in $\ket{\Psi}$.}
   \label{inject}
\end{figure}

As we now have a \cnot operation, a standard procedure can be used to perform these gates on the planar code, given a supply of suitable ancilla states. These ancilla states need to be topologically protected, so we require a state injection procedure for magic states.  The general procedure was given for the planar code in \cite{topo-q-memory}, and a worked example for general surface codes in \cite[\S VI(C)]{austin1}. We here give the exact operations that would be used in a lattice surgery context.

 Figure \ref{inject} demonstrates the procedure for injecting $ \alpha \ket{0} + \beta \ket{1}$. First a distance-3 logical qubit surface is prepared, with all qubits except one data qubit in the \ket{0} state. The remaining qubit is the magic state to be injected. \cnot operations are performed between this qubit and the syndrome qubits immediately above and below it. These syndrome qubits are then swapped with the data qubits immediately above or below to create the 3-qubit state  $\alpha \ket{000} + \beta\ket{111}$, figure \ref{inject}(b). This surface is then stabilized, to give a distance-3 logical qubit in the logical state $ \alpha \ket{0}_L + \beta \ket{1}_L$, figure \ref{inject}(c). Additional physical qubits are prepared in \ket{0}, figure \ref{inject}(d), and then merged into the original logical qubit to give a distance-4 logical qubit in state $ \alpha \ket{0}_L + \beta \ket{1}_L$, figure \ref{inject}(e). The surface may be increased to any desired code distance by further merging.

\subsection{The Hadamard gate}

A useful element in the defect-based surface code is the ability to perform Hadamard gates without needing an Euler decomposition \cite[\S VII]{austin1}. The procedure for creating a Hadamard gate in the planar code differs as there is not a fixed background lattice, and we now give a method for performing this operation.

\begin{figure}[t]
   \centering
    \subfigure[]
       {\includegraphics[width=3.5cm]{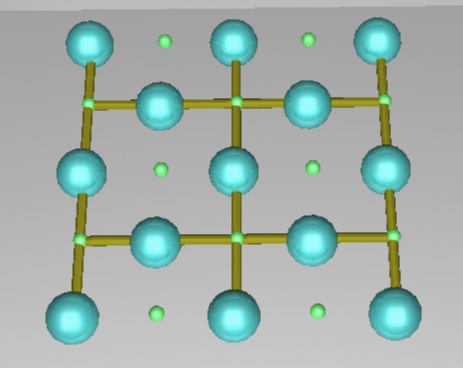}
        }\hspace{1cm}
            \subfigure[]
       {\includegraphics[width=3cm]{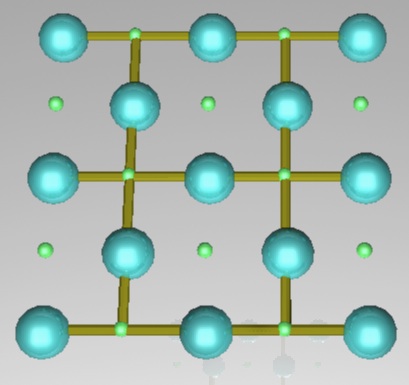}
         }
   \caption{Performing a transversal Hadamard gate leaves the original qubit surface (a) in a rotated orientation (b).}
   \label{had1}
\end{figure}

Performing a Hadamard gate transversally by Hadamard operations on each individual qubit will leave us with a planar qubit that is in the correct state (an eigenstate of $X_L$ is taken to the corresponding eigenstate of $Z_L$ and \emph{vice versa}), but it will leave the planar surface at a different orientation from the original, figure \ref{had1}. If there is no further processing to be done on the qubit then this can remain. Alternatively, if the physical interconnects between planar surfaces are movable, then they can be rotated through 90 degrees on all subsequent interactions. However, if the underlying physical implementation has fixed qubits (or we wish them to remain so for reasons of scalability), then we require a method to rotate the planar surface back to its original orientation.

\begin{figure}[t]
   \centering
    \subfigure[]
       {\includegraphics[width=5cm]{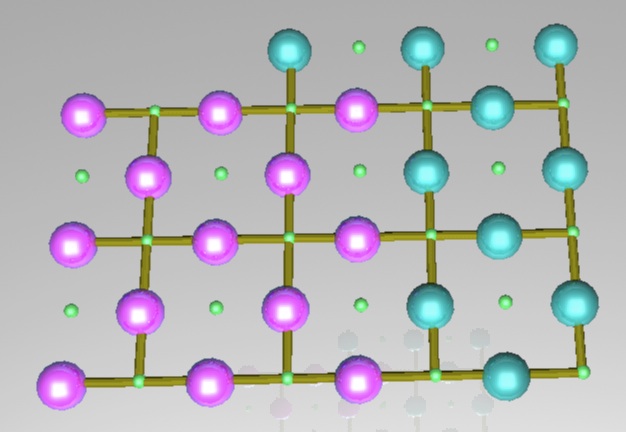}
        }\hspace{1cm}
            \subfigure[]
       {\includegraphics[width=5cm]{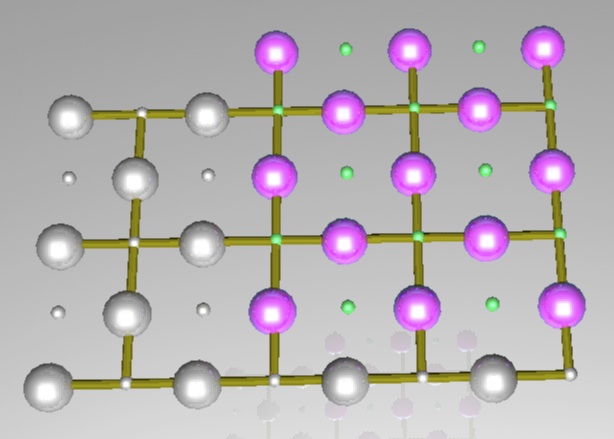}
         }
   \caption{Rotating the orientation of a planar qubit: a) expanding the original surface (pink); b) contracting to form the rotated surface (pink).}
   \label{had2}
\end{figure}

We can perform this rotation using the method shown in figure \ref{had2}. The original surface in figure \ref{had1}(b) is expanded to create the surface in figure \ref{had2}(a). This is still a planar surface with two sets of boundaries, with the smallest logical operator strings between boundaries of length $d$ (the original code distance). Applying $d$ rounds of error correction after this merging maintains fault tolerance. The large surface is then contracted by measuring out qubits in the $Z$ basis, figure \ref{had2}(b). After a further $d$ rounds of error correction this leaves the remaining surface correctly oriented for further interactions with other planar code surfaces, but shifted by half a lattice spacing in both horizontal and vertical directions. This can be corrected by performing swap operations to move the lattice into the correct position.

\section{Relationship to defect qubits}\label{relate}

We saw in \S\ref{scs} that there is a close relationship between logical qubits defined with respect to the boundaries of the lattice, and those defined by the ``extra boundaries" of double defects. We can in fact convert between the two types, extruding a defect-based qubit from the edge of a surface into a planar qubit on a separate surface. As well as demonstrating the connection between the two surface code types, such a procedure could be useful if, for example, single qubits need to be extracted from a larger computation and then distributed. Reducing the number of physical qubits to be communicated in this case would be very useful.


\begin{figure}[t]
   \centering
    \subfigure[]
       {\includegraphics[width=7cm]{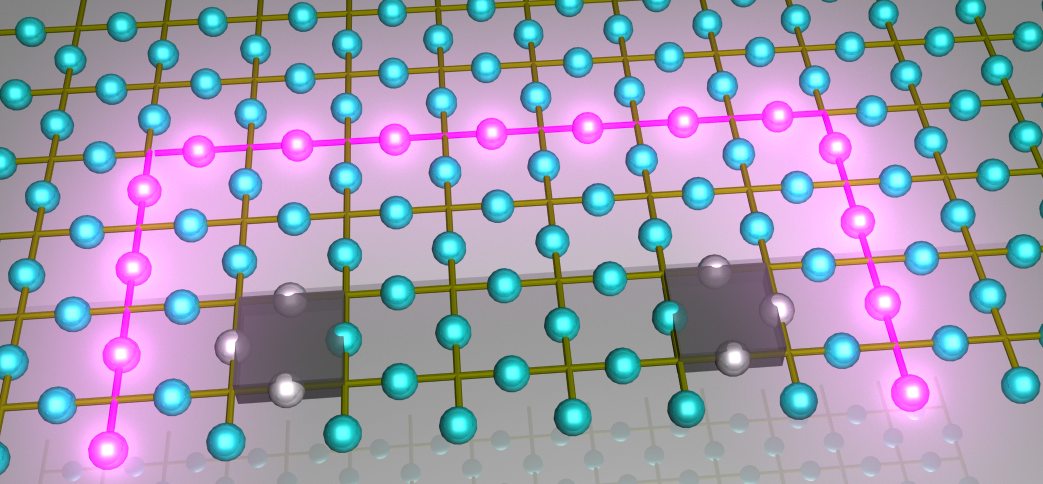}
        }\hspace{1cm}
            \subfigure[]
       {\includegraphics[width=7cm]{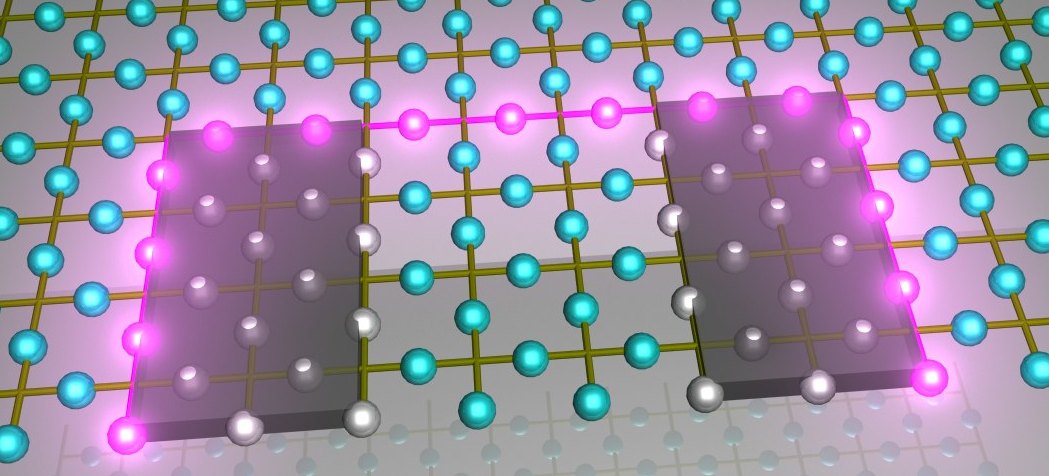}
         }\hspace{1cm}
            \subfigure[]
       {\includegraphics[width=7.5cm]{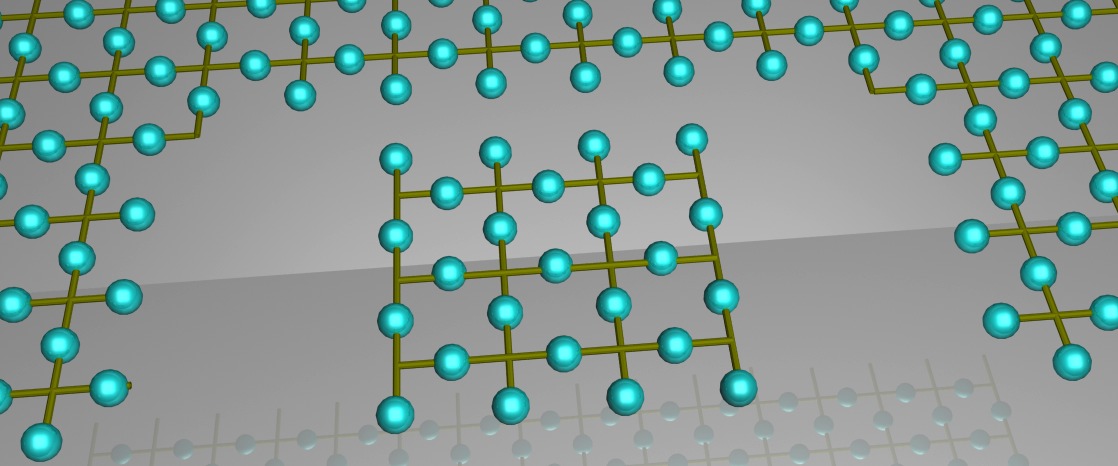}
         }
   \caption{Extracting a planar code qubit from a double defect qubit located at the edge of a larger lattice: a) pink qubits are measured out in the $Z$ basis to isolate the defect qubit from the rest of the lattice; b) the defects are enlarged to fill part of the space; c) unstabilized qubits are removed from the lattice, creating the planar qubit. Note that this procedure also works in reverse.}
   \label{extrude}
\end{figure}

The full procedure, complete with correction operations at the boundary, is detailed in \cite{austinH}. Let us start with a logical qubit defined by a double defect, near the edge of a large lattice, figure \ref{extrude}(a). We can produce a planar qubit from this defect qubit as in figure \ref{extrude}, where the pink qubits are removed from the lattice by measurements. In this case, with a smooth double defect, the measurements are in the $Z$ basis. If the defects were rough, the measurements would need to be in the $X$ basis. This leaves us with an isolated planar qubit, which is no longer attached to the main surface (and can indeed be detached for communication), and encodes the same logical qubit (at the same code distance) as the original defect-based logical qubit.

\section{Resource use of the planar code}\label{resource}

Figure \ref{distfour} shows clearly the significant difference in resources used for planar and defect-based logical codes of the same distance. We can now quantify this difference for individual logical qubits, and also for entangling gates between them. 

For both defect and planar qubits, full fault tolerance requires $d$ rounds of error correction after each operation on a distance $d$ code; operations on a single qubit therefore need the same number of time steps in both cases. However, if we look at the construction of the \cnot operation given in \S\ref{cnot}, it appears that a planar \cnot requires many more rounds of error correction than in a defect-based approach. In fact, if we look closely at the exact operations in the procedure as given, we find that this is not the case. Na\"ively, we would count operations for the steps as: first merge, 3 rounds; split, 3 rounds; second merge, 3 rounds. However, if we look again at the exact procedure as given is \S\ref{cnot}, we find that we do not in fact need the full 9 rounds of error correction. If we prepare the control qubit as a $d \times 2d$ surface, then the first merge operation is not required. Furthermore, all the required operations for the split and the merge commute; they can be performed at the same time, and then three rounds of error correction implemented afterwards. By thus combining the split and second merge, and eliminating the first merge, we can perform the logical \cnot  with only $d$ rounds of error correction; the same as in the defect-based approach. 

As we can make the temporal resources for each implementation equivalent in this way, the figure of merit to consider when comparing defect-based and planar implementations is going to be the number of physical qubits required -- that is, the lattice cross-section or surface area. For a distance $d$ code, a single double-defect qubit is usually implemented using a total lattice area of $\approx 6d^2$ qubits \cite{simon_architecture}. For a planar qubit, the lattice area is $\approx 2d^2$ qubits -- a significant reduction in physical qubits needed.

\begin{figure}[t]
\centering
       \includegraphics[width=4cm]{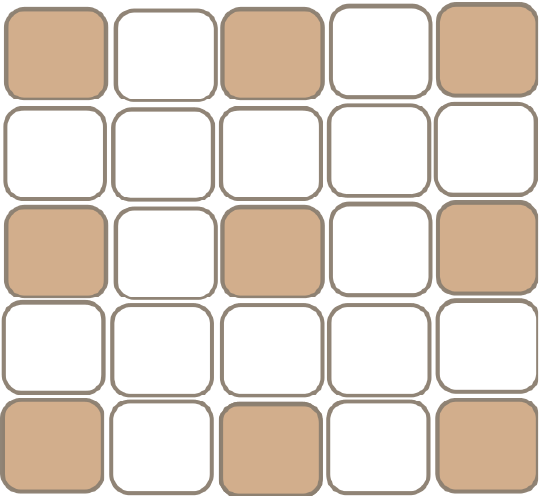}
	\caption{Scalable arrangement of planar code surfaces. Shaded surfaces contain logical data qubits, and blank surfaces are available to perform \cnot operations between data surfaces.}\label{scalableplanar}
\end{figure}

Let us now consider the requirements for the \cnot operation. The double defect implementation uses a large lattice cross-section to allow for the multiple braids that are required. The cross section used is $ \approx 37d^2$ qubits. For the planar code, by contrast, we use three $2d^2$ surfaces, and two lines of $2d$ intermediate data qubits, the equivalent of a single surface of cross section $2d(3d + 2)$. To leading order in $d$, then, this is a total reduction in the number of physical qubits used of around 6 times. We do not claim that the current implementations of the defect-based code are optimal. The straightforward calculation is not, however, the whole story, as we need also in general to consider the layout of qubits that will allow for scalable computation. The numbers given here for the double-defect case are for a layout that is designed to scale arbitrarily. An equivalent layout for planar qubits is shown in figure \ref{scalableplanar}. The ``blank" areas of the lattice are set aside for \cnot operations, in order to allow any qubit to perform a \cnot with any other qubit in the computer. Only a quarter of the available surfaces can then be used to hold logical data qubits -- the rest must remain available for logical operations.  

We can therefore see that, for a large-scale quantum computer, there is very little difference in the resource requirements for planar or defect implementations. The area in which there is a significant difference is where the number of operations and qubits considered is small. For medium-scale processes, with tens of physical qubits, the above comparisons hold. However, if we look at small scale surface code implementations, the appropriate comparison will not be with the double-defect code, but rather the use of a single defect per logical qubit. For the single-defect implementation, a single logical qubit in the best-known implementations uses a surface area of $ \approx 10d^2$ qubits. When considering a single \cnot operation, this can be performed by looping a rough defect (created by not enforcing a vertex stabilizer) all the way around a smooth defect (created by not enforcing a face stabilizer), and uses only the same surface area to leading order in $d$. Exact calculations of the size of the surface depend on the exact code distance and the order of the boundaries used, but the single-defect implementation usually uses between 1.5 and 2 times the qubits of the planar implementation. As we go to smaller scales, and start to make contact with experimentally feasible implementations of the surface code, this difference can become significant.

\section{Small scale experiments on the planar code}

One of the motivations for introducing lattice surgery on the planar surface code was the reduction in qubit requirements for small scale implementations and medium-term achievable experiments. We have seen that the planar implementation indeed uses fewer resources than either the single or double defect model. We now look at exactly how small the planar code can go, and still provide useful experimental results. In order to do this, we first outline a useful modification to the surface code where the qubit lattice is rotated. We then give two proposed medium-term achievable experiments. In the first, lattice surgery is used to create entangled qubits, both Bell pairs and GHZ states. In the second, we give the precise requirements and procedure for the smallest non-transversal planar \cnot. In both cases we find that the addition of lattice surgery to the surface code toolkit significantly reduces the resources required to perform these error correction experiments.

\subsection{The rotated lattice}

The standard method for creating planar encoded qubits uses the square lattice with regular boundaries, as in figure \ref{distfour}(a). However, it is possible to reduce the number of physical qubits required for a single planar surface of a given distance by considering a ``rotated" form of the lattice used. This removes physical qubits from the edges of the lattice, creating irregular boundaries. The shortest string of operators creating a logical operator is frequently then not a straight line; however, as we shall see, it never goes below the code distance of the original surface.
\begin{figure}[t]
   \centering
    \subfigure[]
       {\includegraphics[width=4.5cm]{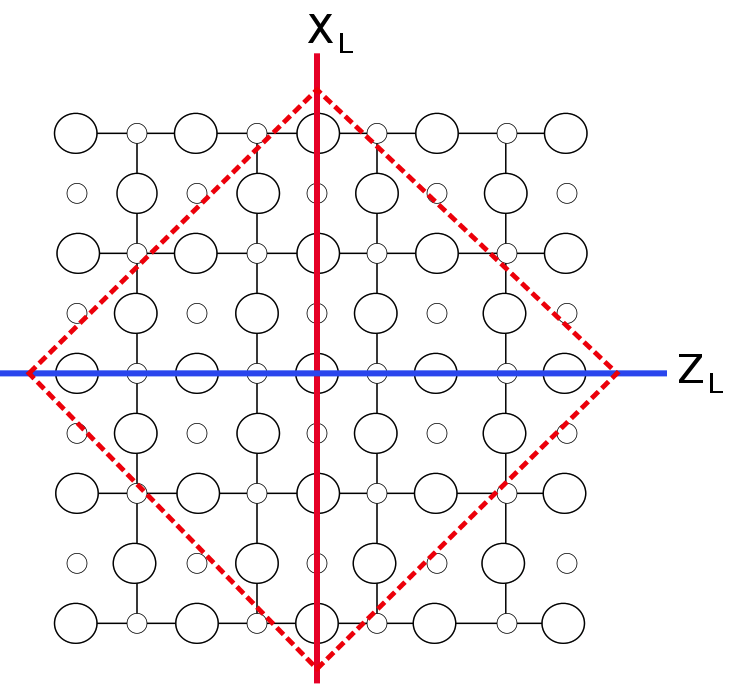}
        }\hspace{1cm}
            \subfigure[]
       {\includegraphics[width=3cm]{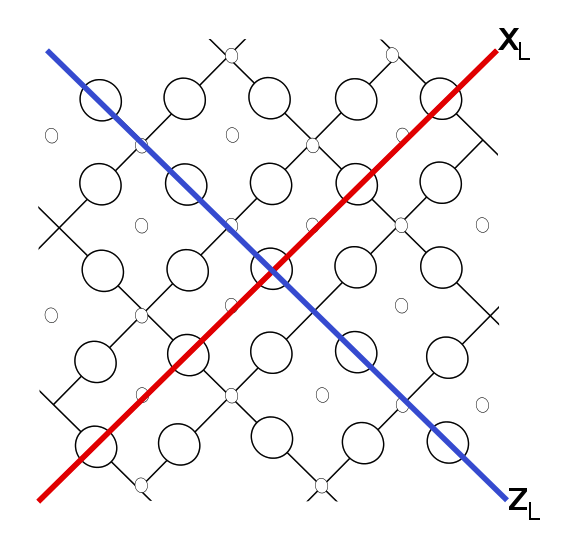}
         }\hspace{1cm}
                   \subfigure[]
       {\includegraphics[width=5.5cm]{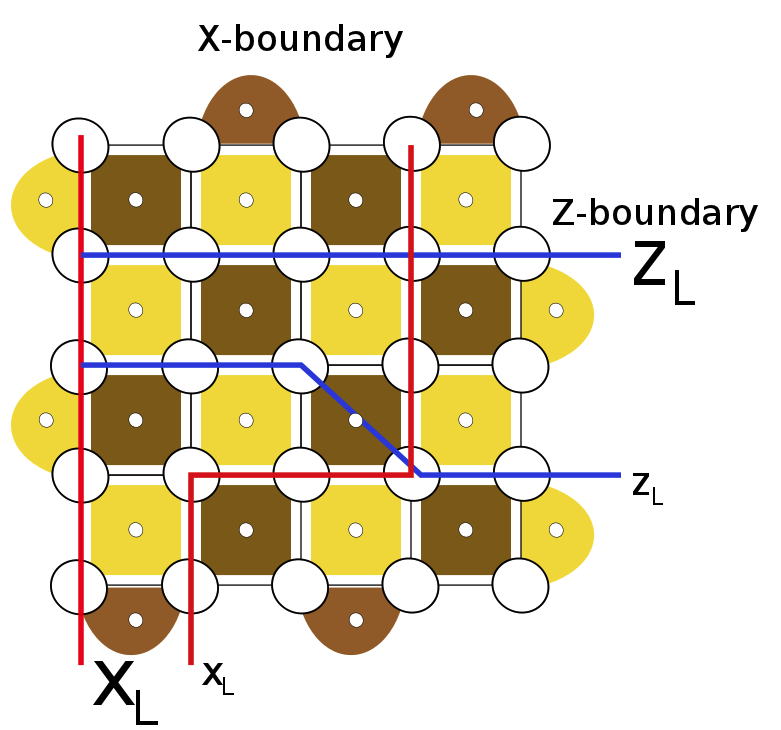}
        }
   \caption{Rotating a distance 5 lattice to produce another distance 5 encoded qubit. a) The original surface. The red square shows the area of the rotated surface. b) The rotated surface; (c) the rotated plaquettes (vertex plaquettes are marked brown, face plaquettes as yellow), new example logical operators, and boundaries. $X$($Z$) logical operators are shown as red(blue) lines.}
   \label{5x5lattices}
\end{figure}

Let us consider the distance 5 code surface shown in figure \ref{5x5lattices}(a). We now create a ``rotated" lattice form by removing all the qubits outside the red box, and rotating (for clarity) $45^\circ$ clockwise. This now has the logical operators and boundaries as shown in figure \ref{5x5lattices}(b). If we now colour in the stabilizers in this new rotated form, we have the lattice in \ref{5x5lattices}(c), with 25 physical qubits and 24 independent stabilizers. The boundaries are now no longer ``rough" or ``smooth". Instead we use \emph{X-boundaries} and \emph{Z-boundaries}: $X$-boundaries have $X$ syndrome measurements along the boundary (shown in the figures as brown), and $Z$-boundaries have Z-syndrome measurements along the edge (shown as yellow)  \cite{planar-bk}. 

There are now more possible paths across the lattice area for a logical operator to take; the smallest paths do, however, stay the same length as before the rotation, so the rotated and unrotated lattices have the same error correction strength. Figure \ref{5x5lattices}(c) shows several example logical operators; note that, in the rotated form, logical operator chain paths can go diagonally through plaquettes of the opposite type (i.e.. $X$-chains can go vertically through $Z$-plaquettes, and \emph{vice versa}). Not all such paths are possible logical operators: care must be taken to ensure that each face(vertex) plaquette is touched an even number of times for an $X$($Z$) operators. This maintains the commutation of the operator with the lattice stabilizers, and the logical operator operations are undetectable by syndrome measurements.  The code distance of the logical qubit remains unchanged, and we have reduced the number of lattice data qubits from $d^2 + (d-1)^2$ to $d^2$ for a distance $d$ code.

\begin{figure}[t]
   \centering
    \subfigure[]
       {\includegraphics[width=4.5cm]{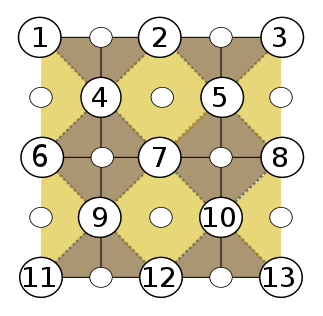}
        }\hspace{1cm}
            \subfigure[]
       {\includegraphics[width=4.5cm]{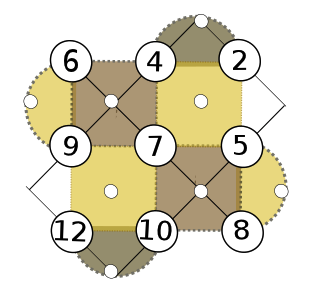}
        }\hspace{1cm}
   \caption{Two lattices encoding a single qubit with distance 3: a) standard planar lattice; b) the rotated lattice. Light(dark) plaquettes show $Z(X)$ syndrome measurements.}
   \label{twolattices}
\end{figure}

The smallest code that detects and corrects one error is the distance 3 code, figure \ref{twolattices}(a). The rotated lattice is shown in figure \ref{twolattices}(b). Note that measuring all the boundary syndromes is unnecessary as the stabilizers are not independent; only the syndromes marked are measured. We can explicitly demonstrate the action of creating the rotated lattice by considering all the stabilizers of the surface, choosing the logical state to be the +1 eigenstate of the $X_L$ logical operator. The left-hand column of figure \ref{stabsrotorig} shows the surface stabilizers of the standard lattice in this case, and the right-hand column gives the corresponding stabilizers in the rotated case. The `rotated' encoding allows us to produce a distance 3 planar qubit with 9 data qubits and 8 syndromes that require measurement. This can be done with either 8 syndrome qubits, or else the four central syndrome qubits can be used twice (while still requiring only neighbouring qubits to interact). The rotated encoding therefore reduces the number physical qubits for the smallest code distance from 25 to 13.
\begin{figure}[t]
\centering
\begin{tabular}{c  || c}
Standard lattice & Rotated lattice\\
stabilizers & stabilizers\\ \hline
$X_2X_7X_{12} \ (=X_L)$ & $X_2X_7X_{12} \ (=X_L)$ \\
$X_1X_2X_4$ & {}\\
$X_2X_3X_4$ & $X_2 X_4$\\
$X_4X_6X_7X_9$ & $X_4X_6X_7X_9$\\
$X_5X_7X_8X_10$ & $X_5X_7X_8X_{10}$ \\
$X_9X_{11}X_{12}$ & {}\\
$X_{10}X_{12}X_{13}$ & $X_{10}X_{12}$\\
$Z_1Z_4Z_6$ & {}\\
$Z_2Z_4Z_5Z_7$ & $Z_2Z_4Z_5Z_7$\\
$Z_3Z_5Z_8$ & $Z_5 Z_8$\\
$Z_6Z_9Z_{11}$ & $Z_6Z_9$\\
$Z_7Z_9Z_{10}Z_{12}$ & $Z_7Z_9Z_{10}Z_{12}$\\
$Z_8Z_{10}Z_{13}$ & {}\\
\end{tabular} 
 \caption{Stabilizers for the distance 3 planar qubit of figure \ref{twolattices}, encoding the state $\ket{+}$, for the standard and rotated lattices. Note that the vertical chain $X_4X_7X_{10}$ is also a valid logical $X_L$ operator. } \label{stabsrotorig}
 \end{figure}
\subsection{Experiment 1: creating entangled states}

We can use lattice surgery, specially the splitting operation, to generate entangled logical Bell pairs and GHZ states. This has an advantage over other methods of preparing entangled logical states that more complicated 2-qubit logical operators are not required. The states created can be n-dimensional GHZ states in either the computational or Hadamard basis.
\begin{figure}[t]
\centering
       \includegraphics[width=12cm]{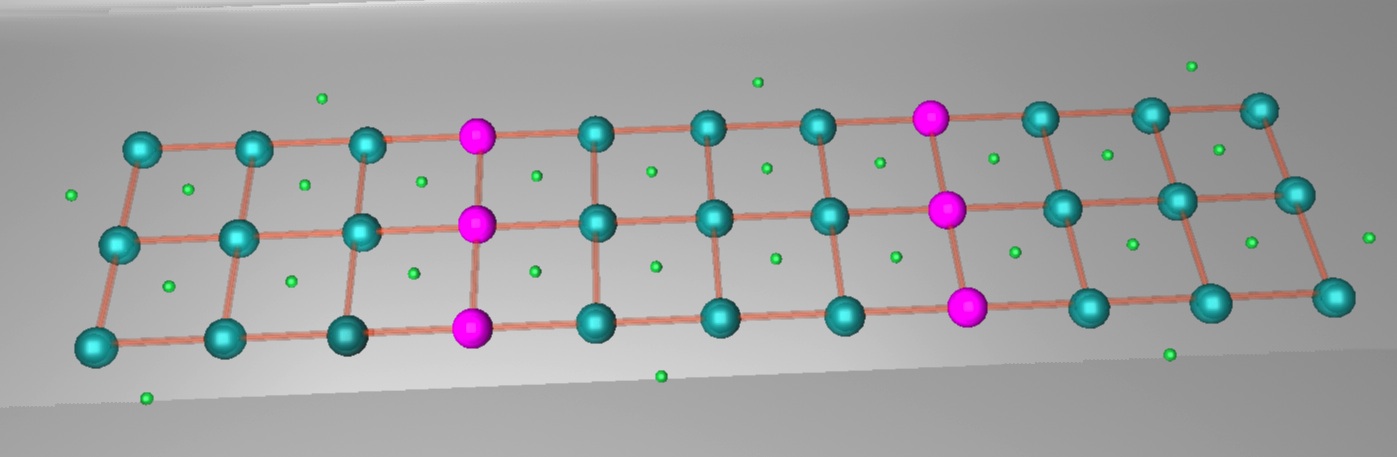}
	\caption{Creating a logical GHZ state on three surfaces in the computational basis through lattice splitting. The orange lattice denotes the rotated encoding.}\label{ghz}
\end{figure}

The procedure is shown in figure \ref{ghz} for distance 4 logical qubits. The initial continuous surface is prepared in the logical $\ket{+}$ state, the +1 eigenstate of the $X_L$ logical operator. The surface is then split by measuring one row of the pink qubits -- this is a smooth split, so the resultant state is
\begin{equation} \ket{+} = \frac{1}{\sqrt{2}}( \ket{0} + \ket{1} ) \longrightarrow \frac{1}{\sqrt{2}}( \ket{00} + \ket{11} )\end{equation}

\noindent which is our Bell pair. If we then measure out the second pink qubit line, we perform another smooth split taking us to a 3-qubit GHZ state:
\begin{equation} \frac{1}{\sqrt{2}}( \ket{00} + \ket{11} ) \longrightarrow \frac{1}{\sqrt{2}}( \ket{000} + \ket{111} )\end{equation}

It is trivial to see that subsequent splittings will produce higher-dimensional GHZ states. The only constraint is that the initial surface is large enough to produce individual logical qubits at the end of the splitting process that retain the desired code distance. We can see that GHZ states in the Hadamard basis can be produced in exactly the same way, with rough splitting. The shape of the original surface will require modification to allow the produced qubits to support the required code distance.

\subsection{Experiment 2: the smallest planar \cnot}

\begin{figure}[t]
\centering
       \includegraphics[width=8cm]{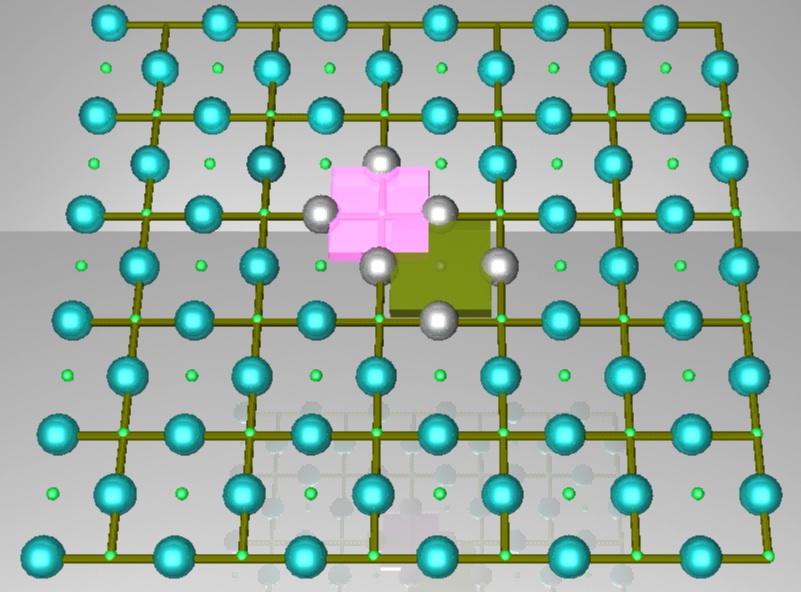}
	\caption{A standard \cnot operation between two logical defects of distance 3 defined by single defects on a lattice. The pink defect is moved around the green one to complete the gate.}\label{cnot-sd}
\end{figure}
The distance 3 rotated planar lattice can also be used to define the smallest number of physical qubits required for a single logical \cnot operation, using the lattice surgery methods of \S\ref{cnot}. With the usual surface code methods, the restriction to nearest-neighbouring-only interactions means that the smallest logical \cnot requires relatively large numbers of physical qubits to implement. Using a double defect, the best-optimised \cnot currently known uses 143 qubits. Using single defects, we have the lattice in figure \ref{cnot-sd}, which also uses 143 physical qubits. By implementing this on a rotated lattice, this can be reduced further to 104 qubits.

\begin{figure}[t]
\centering
       \includegraphics[width=8cm]{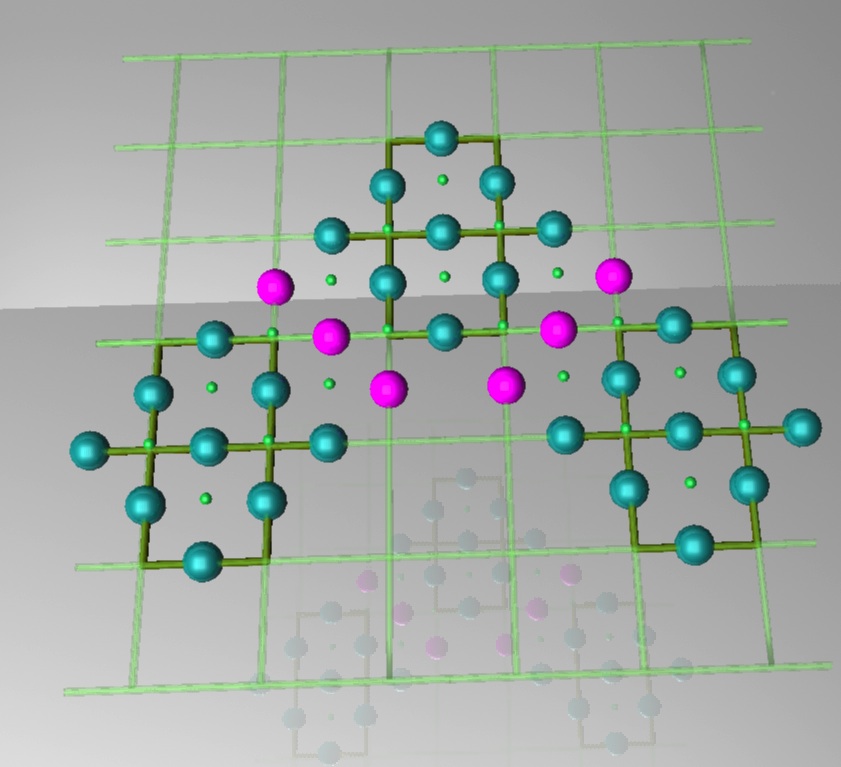}
	\caption{Lattice qubits for a lattice surgery logical \cnot between two distance 3 logical qubits, using the rotated encoding (lattice shown is for the standard encoding to allow comparison with figure \ref{cnot-sd}).}\label{cnot-small}
\end{figure}

The surgery method reduced this number still further. The arrangement of data qubits is shown in figure \ref{cnot-small}. Split and merge operations are performed between the control and target using the method given in \S\ref{cnot}. By using the rotated encoding, and eliminating syndrome qubits outside the data qubit lattice, we require 33 data qubits and 20 syndrome qubits, for a total of 53 physical qubits. We can therefore halve the number of physical qubits required for the smallest \cnot gate implementation by using lattice surgery. While not implementable using current experimental techniques, this number of qubits is by no means out of the question for implementation over the next few years. 

\section{Conclusions}

In this paper we have introduced a method for interacting multiple logical qubits encoded with the planar code, without the need to break the 2DNN structure of the error correction.  This procedure, which we term lattice surgery, allows us to perform coupling operations between encoded qubits without utilising standard transversal protocols.  This method maintains both fault-tolerance and universality while ultimately reducing qubit resources.  

While qubit resource savings are comparatively modest, any saving is advantageous for short to medium term experiments and prototype systems.  The ability of planar codes to be coupled without a full transversal protocol will also be beneficial in the context of distributed computation and/or communication.  As lattice surgery only requires interaction along boundaries, the number of distributed Bell state necessary to perform logical operations between distributed planar qubits is reduced, relaxing the requirements of a repeater network responsible for the connections.  

In addition to the universal set of gates required for large-scale implementation, we have also illustrated some more basic operations: the generation of encoded Bell and GHZ states, and the smallest possible logical operation that can be performed on topologically protected qubits designed to correct for a single arbitrary error on each logical qubit space.  These operations will undoubtably be the first to be experimentally demonstrated once qubit technology reaches the level of 50-100 qubits.  

\section{Acknowledgements}

The authors would like to thank Ashley Stephens and David DiVincenzo for comments on the manuscript. CH, RV and SD acknowledge that this research is supported by the JSPS through its FIRST Program. S.D acknowledges support from MEXT. AGF acknowledges support from the Australian Research Council Centre
of Excellence for Quantum Computation and Communication Technology
(Project number CE110001027), and the US National Security Agency
(NSA) and the Army Research Office (ARO) under contract number
W911NF-08-1-0527.

\section*{References}
\bibliographystyle{unsrt}
\bibliography{SC}

\begin{thebibliography}{10}

\bibitem{threshold}
John Preskill.
\newblock Reliable quantum computers.
\newblock {\em Proc. R. Soc. Lond. A}, 454:385--410, 1998.

\bibitem{topo-q-memory}
Eric Dennis, Alexei Kitaev, Andrew Landahl, and John Preskill.
\newblock Topological quantum memory.
\newblock {\em J. Math. Phys.}, 43:4452, 2002.

\bibitem{raussendorfprl}
R.~Raussendorf and J.~Harrington.
\newblock Fault-tolerant quantum computation with high threshold in two
  dimensions.
\newblock {\em Phys. Rev. Lett.}, 98:190504, 2007.

\bibitem{austin1}
Austin~G. Fowler, Ashley~M. Stephens, and Peter Groszkowski.
\newblock High threshold universal quantum computation on the surface code.
\newblock {\em Phys. Rev. A}, 80:052312, 2009.

\bibitem{austin14}
D.~S. Wang, A.~G. Fowler, and L.~C.~L. Hollenberg.
\newblock Quantum computing with nearest neighbor interactions and error rates
  over 1\%.
\newblock {\em Physical Review A}, 83:020302(R), 2011.

\bibitem{new-threshold-austin}
Austin~G. Fowler, Adam~C. Whiteside, and Lloyd C.~L. Hollenberg.
\newblock Towards practical classical processing for the surface code.
\newblock arXiv:1110.5133, 2011.

\bibitem{bombin}
H.~Bombin and M.A. Martin-Delgado.
\newblock Quantum measurements and gates by code deformation.
\newblock {\em J. Phys. A: Math. Theor.}, 42:095302, 2009.

\bibitem{bomb11}
H.~Bombin.
\newblock Clifford gates by code deformation.
\newblock {\em New J. Phys.}, 13:043005, 2011.

\bibitem{kitaev}
A.~Y. Kitaev.
\newblock Fault-tolerant quantum computation by anyons.
\newblock {\em Annals of Physics}, 303, 2003.

\bibitem{anyonrev}
Gavin~K. Brennen and Jiannis~K. Pachos.
\newblock Why should anyone care about computing with anyons?
\newblock {\em Proc.Roy.Soc. A}, 464(2089):1--24, 1998.

\bibitem{planar-bk}
S.~B. Bravyi and A.~Y. Kitaev.
\newblock Quantum codes on a lattice with boundary.
\newblock quant-ph/9811052, 1998.
\newblock Translation of \emph{Quantum Computers and Computing} 2 (1), pp.
  43-48. (2001).

\bibitem{planar-fm}
Michael~H. Freedman and David~A. Meyer.
\newblock Projective plane and planar quantum codes.
\newblock {\em Foundations of Computational Mathematics}, 1(3):325--332, 2001.

\bibitem{qdos-architecture}
N.~Cody Jones, Rodney~Van Meter, Austin~G. Fowler, Peter~L. McMahon, Jungsang
  Kim, Thaddeus~D. Ladd, and Yoshihisa Yamamoto.
\newblock A layered architecture for quantum computing using quantum dots.
\newblock arXiv:1010.5022v1, 2010.

\bibitem{optics-dots}
David~A. Herrera-Mart\'\i, Austin~G. Fowler, David Jennings, and Terry Rudolph.
\newblock Photonic implementation for the topological cluster-state quantum
  computer.
\newblock {\em Phys. Rev. A}, 82:032332, 2010.

\bibitem{super-archi}
David~P DiVincenzo.
\newblock Fault-tolerant architectures for superconducting qubits.
\newblock {\em Physica Scripta}, T137:014020, 2009.

\bibitem{super-archi-austin}
Peter Groszkowski, Austin~G. Fowler, Felix Motzoi, and Frank~K. Wilhelm.
\newblock Tunable coupling between three qubits as a building block for a
  superconducting quantum computer.
\newblock {\em Phys. Rev. B}, 84:144516, 2011.

\bibitem{atomic-archi}
Jens Kruse, Christian Gierl, Malte Schlosser, and Gerhard Birkl.
\newblock Reconfigurable, site-selective manipulation of atomic quantum systems
  in two-dimensional arrays of dipole traps.
\newblock {\em Phys. Rev. A 81}, 81:060308(R), 2010.

\bibitem{2d-ss-architecture}
Norman~Y. Yao, Liang Jiang, Alexey~V. Gorshkov, Peter~C. Maurer, Geza Giedke,
  J.~Ignacio Cirac, and Mikhail~D. Lukin.
\newblock Scalable architecture for a room temperature solid-state quantum
  information processor.
\newblock arXiv:1012.2864v1, 2010.

\bibitem{2d-rf-iontraps}
Muir Kumph, Michael Brownnutt, and Rainer Blatt.
\newblock Two-dimensional arrays of {R}{F} ion traps with addressable
  interactions.
\newblock {\em New J. Phys.}, 13:073043, 2010.

\bibitem{2d-penning-iontraps}
D.~R. Crick, S.~Donnellan, S.~Ananthamurthy, R.~C. Thompson, and D.~M. Segal.
\newblock Fast shuttling of ions in a scalable {P}enning trap array.
\newblock {\em Rev. Sci. Instrum..}, 81:13111, 2010.

\bibitem{Edmo65a}
J.~Edmonds.
\newblock Paths, trees and flowers.
\newblock {\em Canad. J. Math.}, 17:449--467, 1965.

\bibitem{Edmo65b}
J.~Edmonds.
\newblock Maximum matching and a polyhedron with 0,1-vertices.
\newblock {\em J. Res. Nat. Bur. Standards}, 69B:125--130, 1965.

\bibitem{Kolm09}
V.~Kolmogorov.
\newblock Blossom {V}: a new implementation of a minimum cost perfect matching
  algorithm.
\newblock {\em Mathematical Programming Computation}, 1:43, 2009.

\bibitem{bombtwist}
H.~Bombin.
\newblock Topological order with a twist: Ising anyons from an abelian model.
\newblock {\em Phys.Rev.Lett.}, 105:030403, 2010.

\bibitem{raussendorf3D}
R.~Raussendorf, J.~Harrington, and K.~Goyal.
\newblock Topological fault-tolerance in cluster state quantum computation.
\newblock {\em New J. Phys.}, 9(199), 2007.

\bibitem{bomb06}
H.~Bombin and M.A. Martin-Delgado.
\newblock Topological computation without braiding.
\newblock {\em Phys. Rev. Lett.}, 98:160502, 2007.

\bibitem{algetop}
Joseph Rotman.
\newblock {\em An {I}ntroduction to {A}lgebraic {T}opology}.
\newblock Springer, 1998.

\bibitem{mono}
Andrew Ranicki.
\newblock {\em Algebraic and {G}eometric {S}urgery}.
\newblock Oxford Mathematical Monograph ({O}{U}{P}), 2002.

\bibitem{magic-dist}
Sergey Bravyi and Alexei Kitaev.
\newblock Universal quantum computation with ideal {C}lifford gates and noisy
  ancillas.
\newblock {\em Phys. Rev. A}, 71:022316, 2005.

\bibitem{austinH}
Austin~G. Fowler.
\newblock Low-overhead surface code logical h.
\newblock arXiv:1202.2639, 2012.

\bibitem{simon_architecture}
Simon~J. Devitt, Austin~G. Fowler, Ashley~M. Stephens, Andrew~D. Greentree,
  Lloyd~C.L. Hollenberg, William~J. Munro, and Kae Nemoto.
\newblock Architectural design for a topological cluster state quantum
  computer.
\newblock {\em New Journal of Physics}, 11:083032, 2009.

\end{thebibliography}
\newpage
\appendix
\section{Stabilizer description of merging for a distance-2 code}\label{merge}

\begin{figure}[t]
\centering
       \includegraphics[width=8cm]{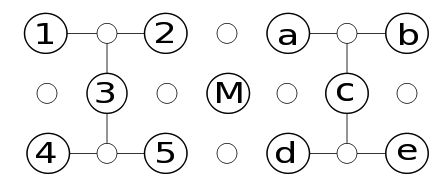}
	\caption{Lattice qubits for merging two rough surfaces of distance 2 into a single surface.}\label{d2merge}
\end{figure}

We give here the full set of stabilizers for the merging operations of \S\ref{lmerge} for two rough surfaces of code distance 2, figure \ref{d2merge}.

The state prior to the merge is $(\alpha \ket{0}_L + \beta\ket{1}_L)\otimes \ket{0} \otimes (\alpha^\prime\ket{0}_L + \beta^\prime\ket{1}_L)$. The stabilizers are therefore a linear combination of 4 sets of terms. The stabilizers for the $\alpha \alpha^\prime \ket{0}_L\otimes\ket{0}\otimes\ket{0}_L$ term are
\begin{equation}
\begin{array}{c c c c c c c c c c c c c c}
{} & {}& 1 & 2 & 3 & 4 & 5 & M & a & b & c & d & e\\
\hline  \hline
\multirow{11}{*}{$\alpha\alpha^\prime$}& {\ldelim \{ {11}{1 mm}} & {X} &X & X & {} & {} & {} & {} & {} & {} & {} & {} & {} \\ 
{} & {}&{} & {} & X & X & X & {} & {} & {} & {} & {} & {} \\ 
{} & {}& Z & {} & Z & Z & {} & {} & {} & {} & {} & {} & {} \\ 
{} & {}& {} & Z & Z & {} & Z & {} & {} & {} & {} & {} & {} \\ 
{} & {}& Z & Z & {} & {} & {} & {} & {} & {} & {} & {} & {} \\ 
{} & {}& {} & {} & {} & {} & {} & Z & {} & {} & {} & {} & {} \\ 
{} & {}& {} & {} & {} & {} & {} & {} & {Z} & Z & {} & {} & {} \\ 
{} & {}& {} & {} & {} & {} & {} & {} & Z & {} & Z & Z & {} \\ 
{} & {}& {} & {} & {} & {} & {} & {} & {} & Z & Z & {} & Z \\ 
{} & {}& {} & {} & {} & {} & {} & {} & X & X & X & {} & {} \\ 
{} & {}& {} & {} & {} & {} & {} & {} & {} & {} & X & X & X \\ 
\end{array}
\end{equation}

The stabilizers $X_2X_MX_a$  are measured across the join to merge the surfaces, with measurement outcome $m$. This element in the state then becomes
\begin{equation}
\begin{array}{c c c c c c c c c c c c c c}
{} & {}& 1 & 2 & 3 & 4 & 5 & M & a & b & c & d & e\\
\hline  \hline
\multirow{11}{*}{$\alpha\alpha^\prime$}& {\ldelim \{ {11}{1 mm}} & {(-1)^m} &  {X} & {} & {} & {} & X & {X} & {} & {} & {} & {} \\ 
{}& {} & X & X & X & {} & {} & {} & {} & {} & {} & {} & {} \\ 
{} & {}&{} & {} & X & X & X & {} & {} & {} & {} & {} & {} \\ 
{} & {}& Z & {} & Z & Z & {} & {} & {} & {} & {} & {} & {} \\ 
{} & {}& {} & Z & Z & {} & Z & {Z} & {} & {} & {} & {} & {} \\ 
{} & {}& Z & Z & {} & {} & {} & {Z} & {} & {} & {} & {} & {} \\ 
{} & {}& {} & {} & {} & {} & {} & {Z} & {Z} & Z & {} & {} & {} \\ 
{} & {}& {} & {} & {} & {} & {} & {Z} & Z & {} & Z & Z & {} \\ 
{} & {}& {} & {} & {} & {} & {} & {} & {} & Z & Z & {} & Z \\ 
{} & {}& {} & {} & {} & {} & {} & {} & X & X & X & {} & {} \\ 
{} & {}& {} & {} & {} & {} & {} & {} & {} & {} & X & X & X \\ 
\end{array}
\end{equation}

\noindent which we can re-write as
\begin{equation}
\begin{array}{c c c c c c c c c c c c c c}
{} & {}& 1 & 2 & 3 & 4 & 5 & M & a & b & c & d & e\\
\hline  \hline
\multirow{11}{*}{$\alpha\alpha^\prime$}& {\ldelim \{ {11}{1 mm}} & {(-1)^m} &  {X} & {} & {} & {} & X & {X} & {} & {} & {} & {} \\ 
{}& {} & X & X & X & {} & {} & {} & {} & {} & {} & {} & {} \\ 
{} & {}&{} & {} & X & X & X & {} & {} & {} & {} & {} & {} \\ 
{} & {}& {} & {} & {} & {} & {} & {} & X & X & X & {} & {} \\ 
{} & {}& {} & {} & {} & {} & {} & {} & {} & {} & X & X & X \\ 
{} & {}& Z & {} & Z & Z & {} & {} & {} & {} & {} & {} & {} \\ 
{} & {}& {} & Z & Z & {} & Z & {Z} & {} & {} & {} & {} & {} \\ 
{} & {}& {} & {} & {} & {} & {} & {Z} & {Z} & Z & {} & {} & {} \\ 
{} & {}& {} & {} & {} & {} & {} & {Z} & Z & {} & Z & Z & {} \\ 
{} & {}& {} & {} & {} & {} & {} & {} & {} & Z & Z & {} & Z \\ 
{} & {}& Z & Z & {} & {} & {} & {} & {Z} & {Z} & {} & {} & {} \\ 
\end{array}
\end{equation}

\noindent The stabilizer $X_5X_MX_d$ across the join is now measured, with outcome $m^\prime$, leaving the state as
\begin{equation}
\begin{array}{c c c c c c c c c c c c c c}
{} & {}& 1 & 2 & 3 & 4 & 5 & M & a & b & c & d & e\\
\hline  \hline
\multirow{11}{*}{$\alpha\alpha^\prime$}& {\ldelim \{ {11}{1 mm}} & {(-1)^m} &  {X} & {} & {} & {} & X & {X} & {} & {} & {} & {} \\ 
{}& {} & (-1)^{m^\prime} & {} & {} & {} & {X} & {X} & {} & {} & {} & {X} & {} \\ 
{}& {} & X & X & X & {} & {} & {} & {} & {} & {} & {} & {} \\ 
{} & {}&{} & {} & X & X & X & {} & {} & {} & {} & {} & {} \\ 
{} & {}& {} & {} & {} & {} & {} & {} & X & X & X & {} & {} \\ 
{} & {}& {} & {} & {} & {} & {} & {} & {} & {} & X & X & X \\ 
{} & {}& Z & {} & Z & Z & {} & {} & {} & {} & {} & {} & {} \\ 
{} & {}& {} & Z & Z & {} & Z & {Z} & {} & {} & {} & {} & {} \\ 
{} & {}& {} & {} & {} & {} & {} & {Z} & Z & {} & Z & Z & {} \\ 
{} & {}& {} & {} & {} & {} & {} & {} & {} & Z & Z & {} & Z \\ 
{} & {}& Z & Z & {} & {} & {} & {} & {Z} & {Z} & {} & {} & {} \\ 
\end{array}
\end{equation}

\noindent where the final line is the logical operator state, and the others give the fully-stabilized new surface. We can write this then in a shorthand form,
\begin{equation}\alpha \alpha^\prime \left\{ \begin{array}{c} [S] \\ Z_1Z_2Z_aZ_b
\end{array} \right.\end{equation}

By performing similar operations on the other components in the state, we find that the merged state of the surface is
\begin{eqnarray} {\hspace{-2cm}} & {} & \alpha \alpha^\prime \left\{ \begin{array}{c} [S] \\ Z_1Z_2Z_aZ_b
\end{array} \right. + \alpha \beta^\prime \left\{ \begin{array}{c} [S] \\ -Z_1Z_2Z_aZ_b
\end{array} \right. + \beta \alpha^\prime \left\{ \begin{array}{c} [S] \\ -Z_1Z_2Z_aZ_b
\end{array} \right. + \beta \beta^\prime \left\{ \begin{array}{c} [S] \\ Z_1Z_2Z_aZ_b
\end{array} \right.\nonumber\\
{\hspace{-2cm}} & = & (\alpha + \beta) \alpha^\prime \left\{ \begin{array}{c} [S] \\ Z_1Z_2Z_aZ_b
\end{array} \right. + (\alpha + \beta) \beta^\prime \left\{ \begin{array}{c} [S] \\ -Z_1Z_2Z_aZ_b
\end{array} \right.\end{eqnarray}

\noindent using the representations given in equation (\ref{mnb}), this is now the stabilizer representation of
\begin{equation} \alpha^\prime (\alpha\ket{0}_L + (-1)^{(m+m^\prime)}\beta\ket{1}_L) + \beta^\prime \sigma_x(\alpha\ket{0}_L + (-1)^{(m+m^\prime)}\beta\ket{1}_L)\end{equation}

\noindent as shown in equation (\ref{stabrep}).

\section{Stabilizer description of splitting for a distance-2 code}\label{split}

\begin{figure}[t]
\centering
       \includegraphics[width=8cm]{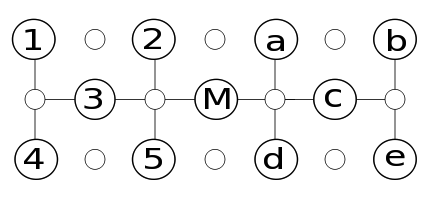}
	\caption{Lattice qubits for splitting a single surface into two distance-2 smooth qubit surfaces).}\label{d2split}
\end{figure}

We give the stabilizer operations for smooth splitting a code surface into two distance-2 planar surfaces as in \S\ref{lsplit}, figure \ref{d2split}.

The stabilizers before splitting are
\begin{equation}
\begin{array}{c c c c c c c c c c c c c c}
{} & {}& 1 & 2 & 3 & 4 & 5 & M & a & b & c & d & e\\
\hline  \hline
\multirow{11}{*}{$\alpha$}& {\ldelim \{ {11}{1 mm}} & {Z} &Z & Z & {} & {} & {} & {} & {} & {} & {} & {} & {} \\ 
{} & {}&{} & {} & Z & Z & Z & {} & {} & {} & {} & {} & {} \\ 
{} & {}& {} & {Z} & {} & {} & {} & {Z} & {Z} & {} & {} & {} & {} \\ 
{} & {}& {} & {} & {} & {} & {Z} & {Z} & {} & {} & {} & {Z} & {} \\ 
{} & {}& {} & {} & {} & {} & {} & {} & Z & Z & Z & {} & {} \\ 
{} & {}& {} & {} & {} & {} & {} & {} & {} & {} & Z & Z & Z \\ 
{} & {}& X & {} & X & X & {} & {} & {} & {} & {} & {} & {} \\ 
{} & {}& {} & X & X & {} & X & {X} & {} & {} & {} & {} & {} \\ 
{} & {}& {} & {} & {} & {} & {} & {X} & X & {} & X & X & {} \\ 
{} & {}& {} & {} & {} & {} & {} & {} & {} & X & X & {} & X \\ 

{} & {}& {} & {Z} & {} & {} & {Z} & {} & {} & {} & {} & {} & {} \\ 
\end{array}
\end{equation}
\begin{equation}
\begin{array}{c c c c c c c c c c c c c c}
{} & {}& 1 & 2 & 3 & 4 & 5 & M & a & b & c & d & e\\
\hline  \hline
\multirow{11}{*}{+$\beta$}& {\ldelim \{ {11}{1 mm}} & {Z} &Z & Z & {} & {} & {} & {} & {} & {} & {} & {} & {} \\ 
{} & {}&{} & {} & Z & Z & Z & {} & {} & {} & {} & {} & {} \\ 
{} & {}& {} & {Z} & {} & {} & {} & {Z} & {Z} & {} & {} & {} & {} \\ 
{} & {}& {} & {} & {} & {} & {Z} & {Z} & {} & {} & {} & {Z} & {} \\ 
{} & {}& {} & {} & {} & {} & {} & {} & Z & Z & Z & {} & {} \\ 
{} & {}& {} & {} & {} & {} & {} & {} & {} & {} & Z & Z & Z \\ 
{} & {}& X & {} & X & X & {} & {} & {} & {} & {} & {} & {} \\ 
{} & {}& {} & X & X & {} & X & {X} & {} & {} & {} & {} & {} \\ 
{} & {}& {} & {} & {} & {} & {} & {X} & X & {} & X & X & {} \\ 
{} & {}& {} & {} & {} & {} & {} & {} & {} & X & X & {} & X \\ 

{} & {}& {} & -{Z} & {} & {} & {Z} & {} & {} & {} & {} & {} & {} \\ 
\end{array}
\end{equation}
If we measure out qubit $M$ in the $X-basis$, with outcome $m$, then the first term becomes
\begin{equation}
\begin{array}{c c c c c c c c c c c c c c}
{} & {}& 1 & 2 & 3 & 4 & 5 & M & a & b & c & d & e\\
\hline  \hline
\multirow{11}{*}{$\alpha$}& {\ldelim \{ {11}{1 mm}}& {(-1)^m} & {} & {} & {} & {} & {X} & {} & {} & {} & {} & {} \\ 
 {} & {} & {Z} &Z & Z & {} & {} & {} & {} & {} & {} & {} & {} & {} \\ 
{} & {}&{} & {} & Z & Z & Z & {} & {} & {} & {} & {} & {} \\ 
{} & {}& {} & {Z} & {} & {} & {Z} & {} & {Z} & {} & {} & {Z} & {} \\ 
{} & {}& {} & {} & {} & {} & {} & {} & Z & Z & Z & {} & {} \\ 
{} & {}& {} & {} & {} & {} & {} & {} & {} & {} & Z & Z & Z \\ 
{} & {}& X & {} & X & X & {} & {} & {} & {} & {} & {} & {} \\ 
{} & {}& {(-1)^m} & X & X & {} & X & {} & {} & {} & {} & {} & {} \\ 
{} & {}& {(-1)^m} & {} & {} & {} & {} & {} & X & {} & X & X & {} \\ 
{} & {}& {} & {} & {} & {} & {} & {} & {} & X & X & {} & X \\ 

{} & {}& {} & {Z} & {} & {} & {Z} & {} & {} & {} & {} & {} & {} \\ 
\end{array}
\end{equation}

\newpage
\noindent which we will write as
\begin{equation}\alpha \left\{ \begin{array}{c} [S_1] \ [S_2] \\ Z_2Z_5Z_aZ_d \\ Z_2Z_5
\end{array} \right.\end{equation}

\noindent where $[S_{1(2)}]$ is the complete set of stabilizers for the first(second) surface after the split. The complete state after the split is then
\begin{equation}\alpha \left\{ \begin{array}{c} [S_1] \ [S_2] \\ Z_2Z_5Z_aZ_d \\ Z_2Z_5
\end{array} \right. + \beta \left\{ \begin{array}{c} [S_1] \ [S_2] \\ Z_2Z_5Z_aZ_d \\ -Z_2Z_5
\end{array} \right.\end{equation}

which is the representation of $\alpha\ket{00}_L + \beta\ket{11}_L$, as in equation (\ref{stabsplit}).
\end{document}